\documentclass[a4paper, 12pt, oneside]{article}

\usepackage{aapreprint}
\usepackage{graphicx,wrapfig,float,slashed,subcaption,bbold,bm}
\usepackage{amsmath,amssymb,epsfig,graphicx,xcolor}
\usepackage{epstopdf}
\usepackage{booktabs}
\usepackage{ragged2e}
\usepackage{mciteplus}

\usepackage{comment}

\usepackage{cancel}

\usepackage{makecell}

\newcommand{\nc}{\newcommand}
\nc{\non}{\nonumber}
\nc{\hc}{\hbox {H.c.}}
\nc{\noi}{\noindent}
\nc{\barx}{\bar{x}}
\nc{\pbarn}{\;\hbox {pb}}
\nc{\fbarn}{\;\hbox {fb}}

\nc{\hsp}{\hspace{0.5cm}}
\nc{\lsp}{\hspace{1cm}}
\nc{\Lsp}{\hspace{2cm}}
\nc{\LLsp}{\lsp\lsp}
\nc{\lra}{\longrightarrow}
\nc{\p}{\prime}
\nc{\sgn}{\text{sgn}}
\nc{\ph}{\varphi}
\nc{\op}{{\cal O}}

\nc{\beq}{\begin{equation}}  \nc{\eeq}{\end{equation}}
\nc{\bea}{\begin{eqnarray}}  \nc{\eea}{\end{eqnarray}}
\nc{\baa}{\begin{array}}     \nc{\eaa}{\end{array}}
\nc{\bit}{\begin{itemize}}   \nc{\eit}{\end{itemize}}
\nc{\ben}{\begin{enumerate}} \nc{\een}{\end{enumerate}}
\nc{\bce}{\begin{center}}    \nc{\ece}{\end{center}}
\nc{\bpm}{\begin{pmatrix}}   \nc{\epm}{\end{pmatrix}}
\nc{\bvt}{\begin{verbatim}}  \nc{\evt}{\end{verbatim}}

\def\lsim{\mathrel{\raise.3ex\hbox{$<$\kern-.75em\lower1ex\hbox{$\sim$}}}}
\def\gsim{\mathrel{\raise.3ex\hbox{$>$\kern-.75em\lower1ex\hbox{$\sim$}}}}

\def\udots{\mathinner{\mkern1mu\raise1pt\vbox{\kern7pt\hbox{.}}\mkern2mu\raise4pt\hbox{.}\mkern2mu\raise7pt\hbox{.}\mkern1mu}}

\def\tev{\;\hbox{TeV}}

\allowdisplaybreaks

\newcommand\fverb{\setbox\fverbbox=\hbox\bgroup\verb}
\newcommand\fverbdo{\egroup\medskip\noindent%
			\fbox{\unhbox\fverbbox}\ }
\newcommand\fverbit{\egroup\item[\fbox{\unhbox\fverbbox}]}
\newbox\fverbbox

\preprint{\begin{flushright}
UTTG 16-2021\\
\end{flushright}}

\title{Twin Quark Dark Matter From Cogenesis} 
\author[a]{Can Kilic,}
\author[b]{Christopher B. Verhaaren,}
\author[a]{and Taewook Youn}
\affiliation[a]{Theory Group, Department of Physics\\ University of Texas at Austin,
Austin, TX 78712, USA}
\affiliation[b]{Department of Physics and Astronomy, Brigham Young University, \\Provo, UT 84602, USA}
\emailAdd{kilic@physics.utexas.edu}
\emailAdd{verhaaren@physics.byu.edu}
\emailAdd{taewook.youn@utexas.edu}

\abstract{
We extend the fraternal twin Higgs scenario to include a novel dark matter candidate as well as a mechanism for generating a matter/antimatter asymmetry in both sectors. A spontaneous breaking of twin color results in quark degrees of freedom that are singlets under the residual twin color group. These twin-color-singlet quarks, along with a subdominant component of twin leptons, constitute the asymmetric dark matter. The asymmetry between matter in antimatter in both sectors is co-generated from the decay of singlet fermions that provide an additional portal between the visible and twin sectors. We discuss the phenomenological aspects of this model, evaluating constraints on the parameter space and highlighting promising discovery channels in future experiments. We briefly discuss how the discovery of signals in multiple experiments may help establish the connection between the mechanisms that address the naturalness, dark matter and matter/antimatter asymmetry puzzles.}

\begin{document}

\maketitle
\flushbottom

%%%%%%%%%%%%%%%%%%%%%%%%%%%%%%%%%%%%%%%%%%%%%%%%%%%%%%%%%%%%%%%%%%%%%%%%%%%%%%%%%%%%%%%%%%%%%%%%%%%%%
\section{Introduction}
\label{sec:intro}

Despite its phenomenal success, the Standard Model (SM) is known to be an incomplete description of fundamental physics. This follows from a number of concerns, which include the naturalness problem of the Higgs potential, the existence of dark matter (DM), and the observed matter/antimatter (M/AM) asymmetry of the Universe. Addressing these issues requires adding new degrees of freedom and interactions to the SM. While it is possible that each of these open questions is explained by entirely disconnected particle sectors, we consider that adding one sector to extend the SM that resolves multiple open puzzles at once offers a more convincing path forward. This work, for example, extends the SM in a way that addresses the Higgs naturalness, dark matter, and M/AM asymmetry puzzles simultaneously. We study the parameter space of this extension in light of existing experimental constraints, as well as discuss future experimental prospects.

Some of the best known extensions of the SM resolve the Higgs naturalness problem by relying on symmetry partner particles that carry the same gauge quantum numbers as their SM partners. However, the null results in searches at the Large Hadron Collider (LHC) for these partner particles have fueled interest in alternative scenarios. Among these, the paradigm of neutral naturalness~\cite{Chacko:2005pe,Barbieri:2005ri,Chacko:2005vw,Burdman:2006tz,Poland:2008ev,Cai:2008au,Craig:2014aea,Craig:2014roa,Batell:2015aha,Csaki:2017jby,Serra:2017poj,Cohen:2018mgv,Cheng:2018gvu,Dillon:2018wye,Xu:2018ofw,Serra:2019omd,Ahmed:2020hiw} has become a very active area of research. In these constructions the partner particles are charged under a separate set of gauge groups than their SM counterparts and are specifically neutral under SM color. Consequently, these color neutral partners are rarely produced at hadron colliders. 

The Twin Higgs (TH) scenario~\cite{Chacko:2005pe} is a simple realization of the neutral naturalness approach, where a twin copy of the SM degrees of freedom and gauge structure is introduced. In addition, the scalar potential has an approximate $SU(4)$ global symmetry, which is spontaneously broken at some scale $f$ producing a pseudo-Nambu-Goldstone Higgs boson. In this set up, only the interactions of the SM and Twin Higgs particles provide a portal between the two sectors. A variant of the TH setup is the Fraternal Twin Higgs (FTH)~\cite{Craig:2015pha}, where the twin sector includes only the third generation of matter fields. While the FTH setup has only a partial twin $Z_2$ symmetry between the two sectors, the largest contributions to Higgs mass quadratic divergences still cancel, and early universe constraints are relaxed because there are fewer light degrees of freedom. In this work we extend the FTH setup, with additional degrees of freedom in both the visible and twin sectors. We also include an addition portal between these sectors in the form of gauge-singlet fermions. This particular portal is somewhat distinct from the various classes of portals that have been already examined in the TH scenario~\cite{Burdman:2014zta,Craig:2015pha,Curtin:2015fna,Ahmed:2017psb,Chacko:2017xpd,Buttazzo:2018qqp,Bishara:2018sgl,Kilic:2018sew,Alipour-Fard:2018rbc,Chacko:2019jgi,Liu:2019ixm,Ahmed:2019kgl}, and could easily exist alongside them.

Since the TH scenario features a rich ``dark sector'' with many degrees of freedom, a number of dark matter candidates have been studied, including twin baryons. While the DM relic abundance is set by thermal freeze-out in some cases~\cite{GarciaGarcia:2015fol,Craig:2015xla,Hochberg:2018vdo,Cheng:2018vaj,Badziak:2019zys,Curtin:2021alk,Curtin:2021spx} (or freeze-in~\cite{Koren:2019iuv}), for others the abundance arises from a M/AM asymmetry in the twin sector~\cite{GarciaGarcia:2015pnn,Farina:2015uea,Barbieri:2016zxn,Barbieri:2017opf,Terning:2019hgj,Beauchesne:2020mih,Chacko:2021vin}. The twin structure relating the visible and hidden sectors certainly suggests the possibility of co-generating a M/AM asymmetry in both sectors simultaneously. In addition, the twin sector already contains degrees of freedom with masses near 5~GeV, so a DM candidate with the same number density as SM baryons appears particularly plausible. This possible connection between DM and asymmetry co-generation between the visible and twin sectors was explored in ref.~\cite{Farina:2016ndq} where the DM is a twin baryon, and a common asymmetry in visible and twin baryon numbers are generated in the early universe through the decay of heavy particles, similar to the mechanism of leptogenesis~\cite{Fukugita:1986hr}. Reference~\cite{Feng:2020urb} also investigate the co-genesis of SM baryons and twin DM. The model we consider is similar to ref.~\cite{Farina:2016ndq} in several aspects, though with significant differences including the structure of the twin sector, in particular the identity of the DM particle.

While not a focus of this work, other connections between Twin Higgs models and cosmology have been explored~\cite{Craig:2016lyx,Chacko:2018vss}. This includes the investigation of cosmological phase transitions~\cite{Schwaller:2015tja,Fujikura:2018duw}, symmetry non-restoration~\cite{Kilic:2015joa}, and high-temperature electroweak symmetry breaking~\cite{Matsedonskyi:2020kuy}. Several scenarios have also been developed to ensure that twin sector contributions to $N_\text{eff}$ agree with experimental limits~\cite{Chacko:2016hvu,Csaki:2017spo,Harigaya:2019shz}. Twin sector effects on large-scale structure~\cite{Prilepina:2016rlq} and the value of the Hubble parameter~\cite{Cyr-Racine:2021alc} have also been studied. 

In a recent paper one of us~\cite{Batell:2020qad} explored the spontaneously breaking of twin color down to an $SU(2)$ residual twin color (RTC) subgroup. After twin color breaking (TCB), in addition to the RTC baryons, some degrees of freedom that originally were colored quarks also become asymptotically free particles. In this work, we consider one such RTC-singlet quark as the dominant DM component, along with a subdominant twin lepton component. In contrast to ref.~\cite{Farina:2016ndq} in which twin color is unbroken, in our model when a M/AM asymmetry is co-generated in the two sectors the RTC-singlet twin top quark acquires an asymmetry, but not the RTC-doublet quarks. As a result, the RTC twin baryons remain symmetric and annihilate efficiently such that their relic abundance today is negligible. The RTC-singlet twin top decays to the RTC twin bottom (the dominant DM component) as well as twin taus and neutrinos through the twin weak interactions.

Our model predicts a variety of experimental signatures. Both the visible and twin sectors contain colored scalars, and those charged under the visible color group can be pair-produced at the LHC and at future hadron colliders, with distinct final states. The LHC does not have sensitivity to the gauge-singlet portal fermions in the parameter space of interest for us, but we do investigate their discovery prospects at future hadron colliders. In the interesting regions of parameter space the presence of the new physics also results in electric dipole moments that exceed the SM prediction. We discuss how discoveries in these channels may provide hints that the solution to the naturalness, DM and M/AM asymmetry puzzles are linked to each other.

Other experimental channels are less sensitive to our model and consequently do not impose significant constraints on the parameter space. Since the DM is asymmetric, it is not possible to observe annihilation signatures in indirect detection experiments. The linking of visible and twin baryon numbers allows the possibility of the RTC-singlet twin bottom decaying into visible sector baryons, however such decays can also easily be forbidden by simple kinematics. Even when the dominant DM component is kinematically allowed to decay, we find that its lifetime is naturally very long, consistent with decaying DM bounds, and the final state of the decay is very challenging to observe. We also show the rate in direct detection experiments and the contributions to flavor-changing processes to be far below the present-day sensitivity. 

In Sec.~\ref{sec:model} we present the particle make-up of our model in full detail. With these new ingredients in hand, in Sec.~\ref{sec:asymmetry} we determine the generation of the M/AM asymmetries in the visible and twin sectors. Following that, in Sec.~\ref{sec:signatures}, we consider all relevant experimental constraints and future prospects, and discuss how future discoveries may point to connections between the solutions to the naturalness, DM and M/AM asymmetry puzzles. We conclude in Sec.~\ref{sec:conclusions}.

%%%%%%%%%%%%%%%%%%%%%%%%%%%

\section{The model}
\label{sec:model}
In this section, we present the field content of our model in quantitative detail. For pedagogical purposes, we separate the Lagrangian into four parts and introduce them one at a time, in the following order: $\mathcal{L}_\text{visible}$, $\mathcal{L}_\text{twin}$, $\mathcal{L}_\text{scalar}$, and $\mathcal{L}_\text{portal}$. 

\subsection{$\mathcal{L}_\text{visible}$}
\label{sec:Lvisible}
The particle content of the visible sector includes the SM fields as well as a new color-triplet scalar $\phi_A$ (we use the label $A$ to signify the visible sector and the label $B$ for the twin sector). The quantum numbers of $\phi_A$ (as well as other new states that we have yet to introduce) are listed in Table~\ref{t.particleQnumbers}. The only new interaction in addition to the SM is a Yukawa coupling between $\phi_A$ and the down-type singlet quarks:
\begin{align}
\mathcal{L}_\text{visible}\supset-Y_LH^\dag_A L_A\overline{E}_A-Y_UQ_AH_A\overline{U}_A-Y_DH^\dag_AQ_A\overline{D}_A -\frac{\lambda}{2}\, \phi_A^\dag\overline{D}_A\overline{D}_A+\text{H.c.}~.\label{e.visLag}
\end{align}
We take all fermion fields to be left-chiral Weyl spinors. We have denoted the $SU(2)_L$ singlet lepton and quark fields as $\overline{E}$, $\overline{U}$, and $\overline{D}$---the bar on top of these fields is just part of the label and does not represent Hermitian conjugation (the latter is expressed with the dagger notation). The first three terms are of course the Yukawa interactions already present in the SM, while the last term introduces the interactions of the new scalar. Note that this term has a color structure proportional to the antisymmetric tensor $\epsilon_{abc}$. Therefore, the flavor indices on the $\overline{D}_{Ai}$ fields need to be antisymmetrized as well, and there are three independent couplings $\lambda_{ij}$.

\begin{table}[th]
\begin{tabular}[t]{l|cccccccccc}
 \toprule 
 & $SU(3)_{A}$ & $SU(2)_{A} $ & $U(1)_{A}$ & $SU(3)_{B}$ & $SU(2)_{B}$ & $U(1)_{B}$ & $B_A$  & $B_B$ & $L_A$ & $L_B$  \\
\midrule
$Q_A$ & $\bm{3}$ & $\bm{2} $ & $\frac16$ & $\bm{1}$ & $\bm{1}$ & 0 & $\frac13$ & 0 & 0 & 0\\
$\overline{U}_A$ & $\bm{\overline{3}}$ & $\bm{1} $ & -$\frac23$ & $\bm{1}$ & $\bm{1}$ & 0 & -$\frac13$ & 0 & 0 & 0\\
$\overline{D}_A$ & $\bm{\overline{3}}$ & $\bm{1} $ & $\frac13$ & $\bm{1}$ & $\bm{1}$ & 0 & -$\frac13$ & 0 & 0 & 0\\
$L_A$ & $\bm{1}$ & $\bm{2}$ & -1 & $\bm{1}$ & $\bm{1}$ & 0 & 0 & 0 & 1 & 0\\
$\overline{E}_A$ & $\bm{1}$ & $\bm{1}$ & 1 & $\bm{1}$ & $\bm{1}$ & 0 & 0 & 0 & -1 & 0\\
$\phi_A$ & $\bm{3}$ & $\bm{1}$ & $\frac23$ & $\bm{1}$ & $\bm{1}$ & 0 & -$\frac23$ & 0 & 0 & 0\\
$\overline{N}_A$ & $\bm{1}$ & $\bm{1}$ & 0 & $\bm{1}$ & $\bm{1}$ & 0 & 1 & 0 & 0 & 0\\
\midrule
$Q_B$ & $\bm{1}$ & $\bm{1} $ & 0 & $\bm{3}$ & $\bm{2}$ & $\frac16$ & 0 & $\frac13$ & 0 & 0\\
$\overline{U}_B$ & $\bm{1}$ & $\bm{1} $ & 0 & $\bm{\overline{3}}$ & $\bm{1}$ &  -$\frac23$ & 0 & -$\frac13$ & 0 & 0\\
$\overline{D}_B$ & $\bm{1}$ & $\bm{1} $ & 0 & $\bm{\overline{3}}$ & $\bm{1}$ & $\frac13$ & 0 & -$\frac13$ & 0 & 0\\
$L_B$ & $\bm{1}$ & $\bm{1}$ & 0 & $\bm{1}$ & $\bm{2}$ & -1 & 0 & 0 & 0 & 1\\
$\overline{E}_B$ & $\bm{1}$ & $\bm{1}$ & 0 & $\bm{1}$ & $\bm{1}$ & 1 & 0 & 0 & 0 & -1\\
$\phi_B$ & $\bm{1}$ & $\bm{1}$ & 0 & $\bm{3}$ & $\bm{1}$ & $\frac23$ & 0 & -$\frac23$ & 0 & 0\\
$N_B$ & $\bm{1}$ & $\bm{1}$ & 0 & $\bm{1}$ & $\bm{1}$ & 0 & 0 & 1 & 0 & 0\\
\bottomrule
 \end{tabular}
\caption{Gauge and global quantum numbers of the relevant matter fields in our model. The first six columns correspond to gauge symmetries in the visible and twin sectors. The following columns correspond to visible and twin baryon and lepton numbers, respectively.
 \label{t.particleQnumbers} }
\end{table}

\subsection{$\mathcal{L}_\text{twin}$}
\label{sec:twin}

Next, we introduce the Lagrangian of the twin (B) sector, which is an extension of the fraternal twin Higgs framework. As such, it contains only one generation of twin matter fields, related by the discrete $Z_2$ twin symmetry ($A\leftrightarrow B$) to the third generation of SM matter fields. Since the $\phi^{\dag}$-$\overline{D}$-$\overline{D}$ interaction is antisymmetric in flavor, it is absent in the twin sector. So in fact, the twin Lagrangian only contains the Yukawa interactions:
\begin{align}
\mathcal{L}_\text{twin}\supset-y_\tau H^\dag_B L_B\overline{E}_B-y_t Q_BH_B\overline{U}_B-y_b H^\dag_BQ_B\overline{D}_B+\text{H.c.}~.
\end{align}
As there is only one generation of twin fermions, the couplings above are simply numbers and not flavor matrices. While we keep the $U/D$ notation in this equation to make the $Z_2$ connection with the visible sector manifest, below we often refer to the twin top and bottom quarks with the symbols $t_B$ and $b_B$ (we use lowercase letters to avoid confusion with the twin baryon number $B_B$).

In the usual FTH scenario the twin spectrum is composed of the twin tau and tau-neutrino along with twin baryons. These baryons are made of three twin $b$-quarks, and they are stabilized by the conserved twin baryon number. Other composite states such as twin glueballs and twin mesons are not stable and they decay to SM states through $H_{A}$-$H_{B}$ mixing, on time scales that are prompt cosmologically, but can be displaced in colliders. 

While a more detailed discussion of the scalar potential is presented in the next subsection, a key feature of our model is that $\phi_B$ acquires a nonzero vacuum expectation value (VEV) $f_{\phi}\sim$~TeV as in ref.~\cite{Batell:2020qad}. We consider here the effects of this on the twin matter fields, by first parameterizing $\phi_{B}$ around its VEV:
\beq
\phi_B=\frac{1}{\sqrt{2}}\left(\begin{array}{c}
0\\
0\\
f_\phi+\varphi_B\\
\end{array} \right)~.
\eeq
This VEV breaks the twin color gauge group from $SU(3)_{c}$ to $SU(2)_{c}$, and five of the twin gluons (as well as the radial mode $\varphi_B$) acquire $f_\phi$ scale masses. As an additional subtlety, because $\phi_{B}$ also carries twin hypercharge, the complete twin sector gauge breaking pattern (when the electroweak symmetry breaking due to the VEV of the twin Higgs is also included) is
\beq
\left[SU(3)_c\times SU(2)_L\times U(1)_Y\to SU(2)_c\times U(1)'_\text{EM} \right]_B,
\eeq
that is, the massless twin photon now also contains part of the twin gluon along the $T^8$ direction. The charge assignments of fields under the massless $U(1)'_\text{EM}$ is given by
\beq
Q_B^\mathrm{'EM}=\sqrt{3}Y_\phi T^8+\tau^3+Y,
\eeq
where we are using the normalization (in the fundamental representation of $SU(3)$ and $SU(2)$), in which
 \beq
T^8=\frac{1}{2\sqrt{3}}\left(\begin{array}{ccc}
1&0&0\\
0&1&0\\
0&0&-2\\
\end{array} \right),
\qquad
\tau^3=\frac12\left(\begin{array}{cc}
1 & 0\\
0 & -1
\end{array} \right),
\eeq
and where $Y_\phi=\frac23$ is the $\phi_B$ hypercharge.

After TCB, a twin quark field $q$ is divided into a color doublet charged under the RTC $SU(2)_c$, which we denote by a hat $\hat{q}_i$ with $i=1,2$, and the RTC singlet $q_3$. When $\phi_B$ gets a VEV, the unbroken twin baryon number becomes the combination:
\beq
B'_B=B_B+\sqrt{3}B_\phi T^8~,
\label{eq:modifiedBB}
\eeq
where $B_\phi=-2/3$ is the baryon number of $\phi_B$. We list the charges of the relevant fields under this unbroken $U(1)$, as well as under the unbroken $U(1)'_\text{EM}$ in Table~\ref{t.baryonNum}. Note that only the RTC singlet fermions carry the unbroken $U(1)$-baryon charge. This means that RTC baryons are not stabilized by this symmetry. They are however stabilized by an accidental global symmetry below the TCB scale (denoted by $T$ in Table~\ref{t.baryonNum}), under which the RTC doublets (and no other fields) are charged.
\begin{table}[th]
\centering
\begin{tabular}[t]{l|ccccccccc}
 \toprule 
 & $\widehat{Q}_{B}$ & $Q_{3B}$ & $\widehat{\overline{U}}_B$ & $\overline{U}_{3B}$ & $\widehat{\overline{D}}_B$ & $\overline{D}_{3B}$ & $\varphi_B$ & $\overline{N}_B$   \\
\midrule
$B_B$ & $\frac13$ & $\frac13 $ & -$\frac13$ & -$\frac13$ & -$\frac13$ & -$\frac13$ & -$\frac23$ & 1 \\
$B'_B$ & 0 & 1 & 0 & -1 & 0 & -1 & 0 & 1 \\
$Q_B^\mathrm{'EM}$ & (1,0) & (0,-1) & -1 & 0 & 0 & 1 & 0 & 0 \\
$T$ & $\frac13$ & 0 & $-\frac13$ & 0 & $-\frac13$ & 0 & 0 & 0 \\
\bottomrule
 \end{tabular}
\caption{Twin baryon number quantum numbers $B_B$ ($B'_B$) before (after) TCB and twin electric charge $Q_B^\mathrm{'EM}$ after TCB. Our notation for the left handed quarks is $Q_B$ = $(U_B,D_B)$. The last line defines the accidental global symmetry $T$ below the TCB scale that stabilizes the RTC baryons.
 \label{t.baryonNum} }
\end{table}

Compared to a Mirror Twin Higgs model, the running of the twin color coupling is modified due to both the fraternal spectrum and the spontaneous color breaking. Within FTH models it is assumed that the SM and twin color couplings are nearly equal at the UV cutoff $\Lambda_\text{UV}\sim$ (few TeV). As shown in~\cite{Craig:2015pha}, the two strong couplings cannot differ by more than about 15\% without introducing additional tuning into the model. Evolving the couplings from $\Lambda_\text{UV}$ towards the IR, the twin coupling initially runs faster than its SM counterpart, because there are fewer light quarks, which in the absence of twin color breaking would lead to a strong couplings scale of a few GeV.

\begin{figure}[t]
	\begin{center}
		\includegraphics[width=0.9\textwidth]{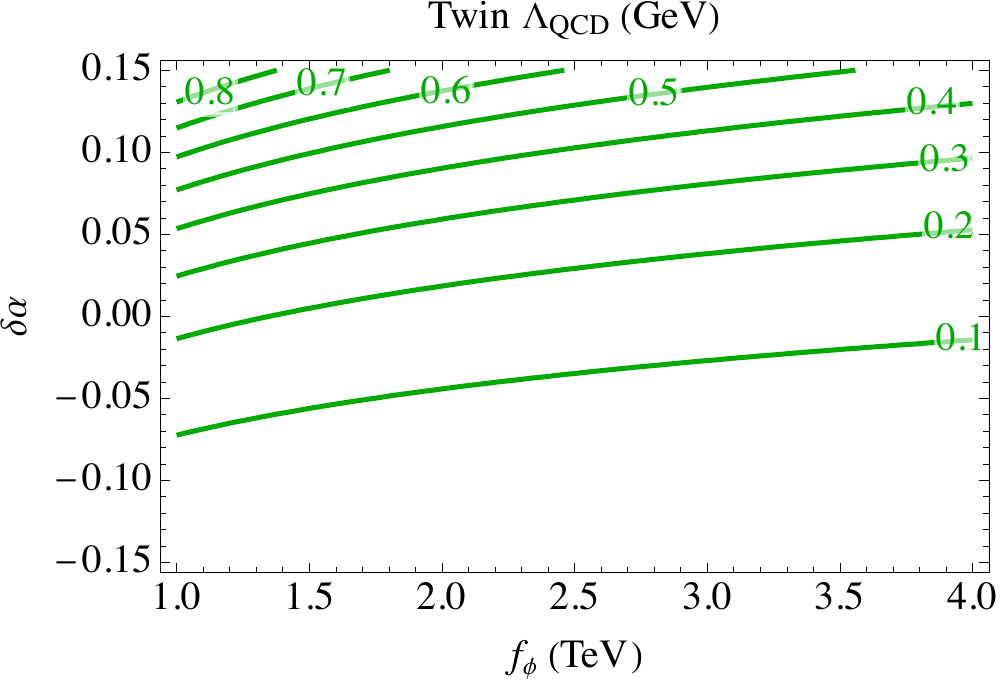}
		\caption{Contours of twin confining scale in GeV as a function of the percent difference $\delta\alpha$ between the twin and SM strong couplings at the scale $\Lambda_\text{UV}$ (here taken to be 5 TeV) and of the TCB scale $f_\phi$. The twin bottom quark is taken to have a mass of 4 GeV and the ratio of the SM Higgs VEV to the $SU(4)$ breaking scale $f$ is taken to be $v/f=1/3$. Variations of these parameter choices do not change the results significantly.}
		\label{fig:twinLam}
	\end{center}
\end{figure}

In the presence of twin color breaking the running slows considerably, due to the change from $SU(3)$ to $SU(2)$ in the beta function. In the Mirror Twin Higgs set up~\cite{Batell:2020qad} (meaning six twin quark flavors and equal couplings at the cutoff) $\Lambda_\text{QCD}$ would be near the MeV scale. Unsurprisingly, a fraternal model with twin color breaking leads to a confining scale in between the MeV and GeV scales. In Fig.~\ref{fig:twinLam} we show contours of the twin $\Lambda_\text{QCD}$ as a function of the percent difference between the SM and twin couplings at $\Lambda_\text{UV}$ and of the TCB scale $f_\phi$. If the visible and twin $\alpha$ are identical at $\Lambda_\text{UV}$ then the twin confinement is near 200 MeV, but it can approach 1 GeV with acceptably non-equal values of $\alpha$ at $\Lambda_\text{UV}$.

The twin bottom quark has a mass of a few GeV, a factor of few above the confining scale. Thus, like heavy quarkonia in the SM, the mesons and baryons (both containing a pair of quarks) can be approximated as nonrelativistic bound states, see~\cite{Kribs:2009fy} for a useful discussion of such objects. Because the fundamental and anti-fundamental representations of $SU(2)$ are interchangeable, the differences between RTC mesons and baryons are less obvious. Bound states with nonzero $T$ number (see Table~\ref{t.baryonNum}) are classified as baryons. As for glueballs, lattice results~\cite{Teper:1998kw,Lucini:2008vi,Athenodorou:2021qvs} indicate that the lightest $SU(2)$ glueball has a mass of $\sim 6.5 \Lambda_\text{QCD}$.

Finally, we assume that the twin photon has a mass. While there are elegant ways to accomplish this, such as including additional degrees of freedom in the Higgs sector, it is sufficient for our purposes to include a Proca mass term for the twin hypercharge boson. As shown in Sec.~\ref{sec:signatures}, a twin photon mass in the neighborhood of a GeV is phenomenologically preferred. The twin photon kinetically mixes with the visible photon through multi-loop effects, but this effect is small and is studied quantitatively in Sec.~\ref{sec:darkphoton}. However, in order to also consider potentially larger values of the mixing, we also allow an explicit mixing term $B^{\mu\nu}B'_{\mu\nu}$. Consequently, we simply treat the mixing parameter $\varepsilon$ as a free parameter, bounded from below by the loop level mixing.

\subsection{${\mathcal L}_{\rm scalar}$}
The scalar potential contains both the usual twin Higgs potential (including $Z_2$ breaking contributions), as well as masses and interactions for the $\phi$'s. The potential features a $Z_2$ symmetry between $\phi_{A}$ and $\phi_{B}$ as in ref.~\cite{Batell:2020qad}:
\begin{align}
{\mathcal L}_{\rm scalar}&=\mu^2\left(H^{\dag}_{A}H_{A}+H^{\dag}_{B}H_{B}\right)+\mu_{\phi}^{2}\left(|\phi_{A}|^2+|\phi_{B}|^2\right)\nonumber\\
&-\lambda\left(H^{\dag}_{A}H_{A}+H^{\dag}_{B}H_{B}\right)^2-\delta\left[\left(H^{\dag}_{A}H_{A}\right)^{2}+\left(H^{\dag}_{B}H_{B}\right)^{2}\right]\nonumber\\
&-\lambda_{\phi}\left(|\phi_{A}|^2+|\phi_{B}|^2\right)^2-\delta_{\phi}\left(|\phi_{A}|^4+|\phi_{B}|^4\right)\nonumber\\
&-\lambda_{H\phi}\left(H^{\dag}_{A}H_{A}+H^{\dag}_{B}H_{B}\right)\left(|\phi_{A}|^2+|\phi_{B}|^2\right)\nonumber\\
&-\delta_{H\phi}\left(H^{\dag}_{A}H_{A}-H^{\dag}_{B}H_{B}\right)\left(|\phi_{A}|^2-|\phi_{B}|^2\right).
\end{align}
The $\lambda$ couplings preserve the global $SU(4)$ and $SU(2)$ symmetries in the Higgs and $\phi$ sectors, respectively. The $\delta$ couplings break the global symmetries, but preserve the twin $Z_2$, $A\leftrightarrow B$. In the case of the Higgs, the $SU(4)$ breaking is assumed to be small so that the pNGB nature of the physical Higgs boson protects its mass from large corrections. The Goldstones of the $\phi$ sector need not be light, so the $\delta_\phi$ coupling can be larger.  

Similar to the analyses of~\cite{Barbieri:2005ri,Batell:2019ptb,Batell:2020qad} when $\delta_\phi<0$ the VEV of the $\phi_{A,B}$ system spontaneously breaks the discrete symmetry, it is either completely in the $A$ sector or completely in the $B$ sector. The phenomenologically viable vacuum preserves SM color, so the VEV is completely in the $B$ sector, breaking twin color. Domain walls related to the breaking of the discrete symmetry do not persist if there is even a very small explicit $Z_2$ breaking term, see~\cite{Batell:2019ptb}.

In the absence of other interactions, and with $\delta>0$ the visible and twin Higgs VEVs would be equal. But when $\phi_B$ acquires its VEV, a $Z_2$ breaking contribution to the Higgs masses results:
\beq
m_{\slashed{Z}_2}^2\left( H^{\dag}_{A}H_{A}+H^{\dag}_{B}H_{B}\right)=\delta_{H\phi}\frac{f_\phi^2}{2}\left( H^{\dag}_{A}H_{A}+H^{\dag}_{B}H_{B}\right)~.
\eeq
As shown in~\cite{Barbieri:2005ri}, this produces a hierarchy between the Higgs VEVs:
\beq
\frac{\langle H_B\rangle^2}{\langle H_A\rangle^2}=\frac{\displaystyle\mu^2\delta+m_{\slashed{Z}_2}^2(2\lambda+\delta)}{\displaystyle\mu^2\delta-m_{\slashed{Z}_2}^2(2\lambda+\delta)}
\eeq
Such a hierarchy is essential, given the LHC limits on Higgs couplings~\cite{Burdman:2014zta}. 

\subsection{${\mathcal L}_{\rm portal}$}
Finally, the portal Lagrangian consists of two Dirac fermions $N_{I=1,2}$ (``portal fermions'') that are complete gauge singlets. We label the left- and right-chiral components of the portal fermions $\overline{N}_{A,\bar I}$ and $N_{B, I}$ respectively, with the former coupling to the visible sector and the latter coupling to the twin sector. The $N_I$ fields have an approximate $SU(2)_{\overline{N}_A}\times SU(2)_{N_B}$ flavor symmetry, which is broken by their couplings to the visible and twin sector fields, as shown below. Suppressing flavor indices, the portal Lagrangian is
 \beq
 \mathcal{L}_\text{portal}\supset-M_N\overline{N}_A N_B-\kappa_A\phi_A\overline{U}_A\overline{N}_A-\kappa_B\phi_B\overline{U}_B N_B+\text{H.c.}~,
 \label{eq:Lportal}
 \eeq
and preserves the fraternal $Z_2$ with $\overline{N}_A\leftrightarrow N_B$. Note that once we expand in quark and $N$ flavors, there are eight independent (complex) couplings: six $\kappa_{A,i\bar I}$, and two $\kappa_{B,J}$. Despite the choice of name, the $N$'s are not right handed neutrinos. Their nonzero baryon number forbids any interactions of the form $HLN$ in either the visible or twin sectors.
 
We take the masses of the two $N$ flavors to be nearly equal in the UV, with only a small fractional splitting $\xi$. This can be accomplished by an approximate $SU(2)_{\overline{N}_A}\times SU(2)_{N_B}$ flavor symmetry in the UV, with a scalar bilinear with Yukawa coupling to $\overline{N}_A$ and $N_B$ acquiring a VEV close to the identity. Furthermore, in the IR an additional mass splitting is induced by the $\kappa$ couplings, which act as spurions of the $N$ flavor symmetry. More precisely, one-loop effects give different wave function normalizations to the portal fermions. Canonically normalizing their kinetic terms results in a fractional shift in their masses. These effects can be summarized as follows:
\beq
\left(M_N\right)_{\bar{I}J}=M_{0}\left(\delta_{\bar{I}J}+\xi\ \sigma^{3}_{\bar{I}J}\right)+\frac{c_\Delta M_0}{16\pi^2}\left(\sum_{i}\kappa_{A,i\bar{I}}\kappa^{*}_{A,iJ}+\kappa^{*}_{B,\bar{I}}\kappa_{B,J}\right),
\label{eq:Nmasses}
\eeq
with $\sigma^3$ being the third Pauli matrix, and $c_\Delta$ an order one number.

Note that the $N$ mass term breaks the individual baryon number of the $A$ and $B$ sectors. However, it preserves the combination $U(1)_{B_A-B_B}$, or more precisely $U(1)_{B_A-B'_B}$ after twin color breaking. This mass term includes mixing between the portal fermions. In what follows we assume that the $2\times2$ mass matrix above has been diagonalized, and the $\kappa$ couplings are defined in the basis where this is true. We refer to the mass eigenstates of the $2\times2$ mass matrix as $M_{\pm}$. 
 
Twin color breaking has a number of significant effects on the twin sector. In the limit where the visible and twin baryon numbers are separately conserved, the RTC singlet quarks are asymptotic states and are stable, and the same is true of RTC baryons. However, the visible and twin baryon numbers are not separately conserved but broken down to $U(1)_{B_A-B'_B}$ due to $N$ mass terms, which allows  RTC singlet quarks to decay to SM states, if this is kinematically allowed. The RTC baryons, on the other hand, remain stable due to the accidental symmetry denoted by $T$ in Table~\ref{t.baryonNum}, as already mentioned. The RTC singlet bottom $b_{3B}$ is the dominant DM component in our model. If it is kinematically allowed to decay, it is therefore classified as decaying DM. We estimate its lifetime in Sec.~\ref{sec:decayingDM} to evaluate the corresponding constraint on the model parameters, but there is also a region of parameter space where $b_{3B}$ is stable. 
 
 Another effect of TCB is to allow the RTC singlet top to mix with the portal fermions. As shown in the next section, this plays a significant role in the generation of the baryon asymmetry in the twin sector, and therefore we study the mixing quantitatively below. Also keeping the twin electroweak VEVs, we start with
 \beq
 \mathcal{L}_\text{mass}\supset-\frac{v_By_t}{\sqrt{2}}u_{3B}\overline{U}_{3B}-\frac{\kappa_Bf_\phi}{\sqrt{2}}\overline{U}_{3B}N_{B}-M_N \overline{N}_{A} N_{B}+\text{H.c.}~.
 \eeq
These terms can be written in $3\times 3$ matrix form
 \beq
  \left(\overline{N}_{A,1},\,\overline{N}_{A,2},\, \overline{U}_{3B}\right)\frac{1}{\sqrt{2}}\left(\begin{array}{ccc}

\sqrt{2}M_+ & 0 & 0\\
0 & \sqrt{2}M_- & 0 \\
 \kappa_{B,1}f_\phi & \kappa_{B,2}f_\phi &   y_t v_B  \\
 \end{array} \right)\left(\begin{array}{c}
 N_{B,1}\\
 N_{B,2}\\
   u_{3B}
 \end{array} \right).
 \eeq
Let us define the matrix above as $\mathcal{M}_{F}$, and diagonalize it by way of the unitary matrices $U$ and $V$
 \beq
 U^\dag
 \mathcal{M}_F
 V
 =\left(\begin{array}{ccc}
M_{n_{+}}  & 0 & 0 \\
 0 & M_{n_{-}}  & 0\\
 0 & 0 & M_{t_{3B}}
 \end{array} \right)~,
 \eeq
where $M_{t_{3B}}$ is the RTC-singlet  top mass eigenvalue, and $M_{n_{\pm}}$ are the masses of the two orthogonal portal fermion mass eigenstates. The mass eigenstates are then identified as
 \beq
\left(\overline{N}_{A,1},\;\overline{N}_{A,2},\;\overline{U}_{3B} \right)  = \left(\overline{n}_{+},\;\overline{n}_{-},\;\overline{t}_{3B} \right) U^\dag, \ \ \ \  
\left(\begin{array}{c}
 N_{B,\bar{1}}\\
 N_{B,\bar{2}}\\
   U_{3B}
 \end{array} \right) 
 =V \left(\begin{array}{c}
 n_{+}\\
 n_{-}\\
  t_{3B}
 \end{array} \right) .\label{e.massEigs}
 \eeq
 This motivates the definitions
 \begin{align}
 \kappa_{A+}&\equiv\kappa_{A1}U^\ast_{1,1}+\kappa_{A2}U^\ast_{2,1}~,& \kappa_{A-}&\equiv\kappa_{A1}U^\ast_{1,2}+\kappa_{A2}U^\ast_{2,2} ~,& \kappa_{At}&\equiv\kappa_{A1}U^\ast_{1,3}+\kappa_{A2}U^\ast_{2,3}~,& \nonumber\\
 \kappa_{B+}&\equiv\kappa_{B1}V_{1,1}+\kappa_{B2}V_{2,1} ~,& \kappa_{B-}&\equiv\kappa_{B1}V_{1,2}+\kappa_{B2}V_{2,2} ~, & \kappa_{Bt}&\equiv\kappa_{B1}V_{1,3}+\kappa_{B2}V_{2,3}~.
 \end{align}
 We can then express the portal interactions in terms of the elements of the $U$ and $V$ matrices, which informs our discussion of the twin baryon asymmetry in the next section:
\begin{align}
 &-\phi_A\overline{U}_A\left(\overline{n}_+\kappa_{A+}+\overline{n}_-\kappa_{A-}+\overline{t}_{3B}\kappa_{At}\right)+\text{H.c.}\\
 & -\frac{\varphi_B}{\sqrt{2}}\left(\overline{n}_+U^\ast_{3,1}+\overline{n}_-U^\ast_{3,2}+\overline{t}_{3B}U^\ast_{3,3} \right)
\left(n_+\kappa_{B+}+n_-\kappa_{B-}+t_{3B}\kappa_{Bt} \right)+\text{H.c.}~.\label{e.MassBasisPortal}\nonumber
 \end{align}

\subsection{Simplified description of the parameter space}
\label{sec:simplified}
Our model has many input parameters. In order to make the quantitative analyses of the rest of the paper easier to follow, we now introduce a simplified set of parameters, which are sufficient for a representative discussion of the phenomenology. We take the three independent couplings $\lambda_{ij}$ of the $\phi_A$-$\overline{D}$-$\overline{D}$ interaction of Eq.~\eqref{e.visLag} to be similar in magnitude, and use $\lambda$ to stand for all of them. Similarly, we use $\kappa$ to stand in for all $\kappa_{(A,B)(+,-,t)}$. 

For certain aspects of the phenomenological discussion, small differences between the various $\lambda$ and $\kappa$ couplings have no significant impact. In these cases we take them to be exactly equal when plotting constraints etc. When discussing other features of the phenomenology, such as the generation of the M/AM asymmetry, the various entries of the $\kappa$ couplings being not exactly equal to each other is crucial. In those cases, we conduct Monte Carlo studies, randomly assigning these entries with a similar magnitude and random phases, and we keep track of the median values of quantities of interest. 

With these simplifications, most phenomenological results can be summarized by using the $\lambda$-$\kappa$ notation. Apart from these, the only other parameters of note are the photon-twin photon mixing parameter $\varepsilon$ and the masses, which are scanned to describe certain aspects of the phenomenology, and set to a benchmark value for others. The details of the scalar sector parameters do not play a significant role in the rest of the paper apart from ensuring that $\phi_{B}$ gets a TeV scale VEV.

%%%%%%%%%%%%%%%%%%%%%%%%%%%
\section{Baryogenesis and DM asymmetry generation}
\label{sec:asymmetry}

In this section we discuss how the M/AM asymmetry is generated in the visible and twin sectors. The mechanism is similar to that of ref.~\cite{Farina:2016ndq}, but with a few important differences. The portal fermions $N$ are produced non-thermally when the universe reheats after inflation, and their out-of-equilibrium decays populate both the visible and twin sectors. Note that the $\kappa_{A,iI}$ and $\kappa_{B_I}$ couplings in the portal sector contain physical phases that source CP violation necessary to generate an asymmetry. The diagonal baryon number $B_A-B'_B$ is conserved in our model as described in the previous section, so no {\it net} asymmetry can be generated, but that is no obstacle to equal baryon number densities being generated in the visible and twin sectors. 

As described below, when the $N$'s decay through the portal interaction, the asymmetry is generated first in the RTC singlet twin tops $t_{3B}$. However, these decay quickly through the twin weak interactions, so the asymmetry is transferred to $b_{3B}$, $\tau_{B}$ and $\nu_B$. Therefore, in terms of the asymmetric matter content in the universe, for each visible baryon, the twin sector contains one RTC singlet twin bottom, one twin (anti)tau, and one twin neutrino. We take the twin neutrinos to be light enough to be treated as massless. Cosmological problems associated with this choice can be evaded by taking the temperature of the twin sector to be lower than the visible sector, which in turn can be accomplished by the portal fermions to have a slightly lower branching ratio into the twin sector than the visible one, similar to~\cite{Chacko:2016hvu}. Note that this does not interfere with equal asymmetries being generated in the two sectors, which is guaranteed by the conserved $B_A-B'_B$ number symmetry.

We can turn a knowledge of $\Omega_{\rm DM}$ into a statement about the twin bottom and tau masses: 
\beq
\frac{\Omega_\text{DM}}{\Omega_B}=\frac{m_{b_{3B}}+m_{\tau_B}}{m_p}~,
\eeq
where $m_p$ is the proton mass. Using the cosmological parameters given in ref.~\cite{Aghanim:2018eyx}, we get
\beq
m_{b_{3B}}+m_{\tau_B}=m_p\frac{\Omega_\text{DM}}{\Omega_B}= 4.99\pm0.05\,\text{GeV}~.
\label{eq:mx}
\eeq
Here, the ratio of the twin bottom and tau masses may be the same as the ratio of the visible bottom and tau masses, but does not have to be. In Sec.~\ref{sec:decayingDM} we describe how the choice of these masses may keep the twin bottom absolutely stable, or allow it to decay over extremely long timescales.

If the asymmetries in the two sectors were to be generated above the TCB scale $f_\phi\sim$ TeV, they would wash each other out by processes of the form $\phi_Aq_A\leftrightarrow\phi^\ast_Bq^\dag_B$. Therefore, we consider asymmetry generation at temperatures $T\lsim m_{\phi}/25$. With $m_\phi\sim f_\phi\sim$~TeV, this means that the asymmetry is generated at $T\sim \mathcal{O}(10~{\rm GeV})$. With the reheat temperature thus being below the electroweak scale, sphalerons in the visible sector are not effective in generating a lepton asymmetry. However, while no {\it net} lepton number is generated, charge conservation ensures that an equal number of charged leptons and antineutrinos are created through the weak interactions (which do not decouple until $T\sim \mathcal{O}(10~{\rm MeV})$) to offset the net charge of the protons such that the universe remains charge-neutral. 

One important consequence of the asymmetry being generated below TCB is that it is generated in the modified twin baryon number $B'_B$ of Eq.~\eqref{eq:modifiedBB}. As listed in Table~\ref{t.baryonNum}, only RTC-singlet quarks carry $B'_B$ number, but not the RTC-doublet quarks or twin baryons. Therefore, the asymmetry in the twin sector is generated only in the RTC-singlet quarks, more specifically the RTC-singlet top, which then decays quickly through the twin weak interactions.

Before turning our attention to a quantitative analysis of the asymmetry generation, we give a quick summary of the thermal history. As already mentioned, both sectors are populated through out-of-equilibrium decays of the portal fermions which alone are produced in reheating. By making the reheaton lifetime long, the number density of these initial portal fermions can be controlled, which, once their decay products thermalize, sets the reheat temperature. In other words, there is no contradiction with reheating starting with (out-of-equilibrium) particles whose masses are larger than the reheat temperature. 

As the portal fermions decay, $\phi$'s and up-type quarks are produced first from the portal interactions, and then these continue decaying and populating the lighter species until a thermal distribution is reached. The time scales for all annihilation and decay processes can be shown to be fast enough for this to happen. It is already known that a reheat temperature of $\mathcal{O}(10~{\rm MeV})$ is viable for the visible sector. In the twin sector, all degrees of freedom with a mass above a GeV are either RTC colored, or charged under $Q_B^\mathrm{'EM}$, or have two-body weak decays, therefore they annihilate to RTC gluons, twin photons, or decay through the weak interactions. The annihilation of charged particles to twin photons is efficient even though the twin photons have a nonzero mass, as long as it is kinematically allowed. These degrees of freedom then efficiently thermalize with the SM as long as the twin photons decay sufficiently fast to pairs of SM fermions through kinetic mixing. In order for the symmetric component of twin taus and bottoms to annihilate efficiently and leave behind only the asymmetric component, we require that both be heavier than the twin photon.

 \begin{figure}
	\begin{center}
		\includegraphics[width=1\textwidth]{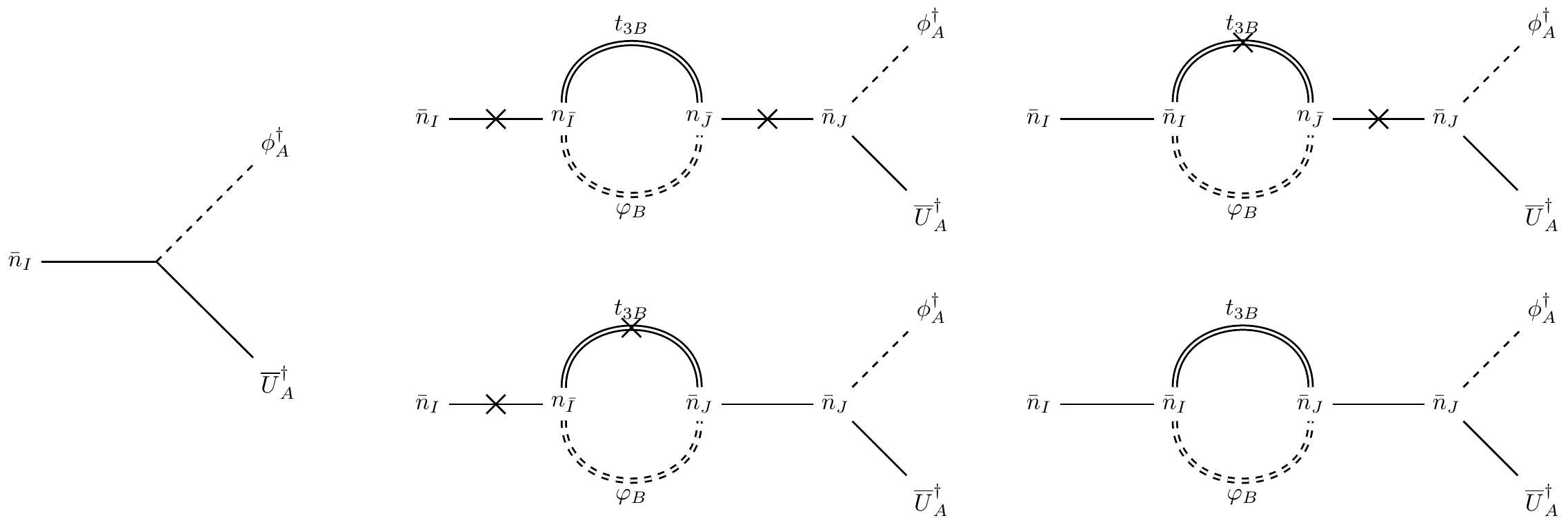}
		\caption{Feynman diagrams relevant to generating a M/AM asymmetry in the visible sector.}
		\label{fig:asym}
	\end{center}
\end{figure}

Having summarized the thermal history of the twin sector, we proceed to calculate the size of the asymmetry generated by $N$ decays. We work with comoving yields $Y_f=n_f/s$ for a given particle $f$. The baryon asymmetries in either sector come from the out of equilibrium decays of the portal fermions 
\beq
Y_{B_A}=Y_{B'_B}=\eta Y_N~,
\eeq
where the asymmetry parameter $\eta$ is a product of two factors: 
\beq
\eta=\left(\epsilon_{A+}+\epsilon_{A-}\right)\times W
\label{eq:defineeta}
\eeq
The first factor has to do with the generation of the asymmetry from the decays of the portal fermions, and it arises through the interference between the tree level and one-loop diagrams of Fig.~\ref{fig:asym}. The second factor accounts for a potential suppression of the asymmetry due to washout processes, and is discussed below.  The asymmetry generated in the twin sector is guaranteed to be equal to the one generated in the visible sector due to the unbroken $B_A-B'_B$ symmetry. Therefore, below we only present the calculation for the asymmetry in the visible sector, but we have verified that the explicit calculation of the asymmetry in the twin sector gives an identical result.

The $\epsilon_{A\pm}$ can be calculated by performing a sum over all final states $X$ with baryon number $B_A(X)$, arising from the decays of the portal mass eigenstates.
\beq
\epsilon_{A\pm}=\sum_XB_A(X)\left[\text{BR}\left(n_\pm\to X_A \right)-\text{BR}\left(n^\dag_\pm\to X^\dag_A \right) \right].
\eeq
Explicitly, this sum can be written as
\beq
\epsilon_{A\pm}=\frac{\Gamma\left(n_\pm\to \overline{U}^\dag_A\phi_A^\dag \right)-\Gamma\left(n^\dag_\pm\to \overline{U}_A\phi_A \right)}{\Gamma\left(n_\pm\to \overline{U}^\dag_A\phi_A^\dag \right)+\Gamma\left(n_\pm\to t^\dag_{3B}\varphi_B  \right)+\Gamma\left(n_\pm\to \overline{t}^\dag_{3B}\varphi_B  \right)}~.
\eeq

The diagrams of Fig.~\ref{fig:asym} contribute different coupling combinations to $\epsilon_{A\pm}$. The leading results, up to $\mathcal{O}(m_\phi^2/M_{n_+}^2)$, are
\begin{align}
\epsilon_{A+}=\epsilon_{B+}\approx&~ \mathcal{R}\times\frac{M_{n_-}}{4\pi M_{n_+}}\left(\frac{\text{Im}\left\{\kappa_{A+}\kappa^\ast_{A-}\left[U_{3,1}U_{3,2}^\ast|\kappa_{Bt}|^2+|U_{3,3}|^2\kappa_{B+} \kappa_{B-}^\ast \right] \right\}}{2|\kappa_{A+}|^2+|U_{3,1}^\ast\kappa^\ast_{Bt} |^2+|U_{3,3}^\ast\kappa_{B+}|^2}\right.
\nonumber\\
&+2\left.\frac{M_{t_{3B}}}{M_{n_+}} \frac{\text{Im}\left\{\kappa_{A+}\kappa^\ast_{A-}\left[U^\ast_{3,3}U^\ast_{3,2}\kappa_{B+}\kappa_{Bt}+U_{3,2}U_{3,3}\kappa_{B-}^\ast\kappa_{Bt}^\ast  \right] \right\}}{2|\kappa_{A+}|^2+|U_{3,1}^\ast\kappa^\ast_{Bt} |^2+|U_{3,3}^\ast\kappa_{B+}|^2}\right)~,
 \label{eq:epsp}\\
\epsilon_{A-}=\epsilon_{B-}\approx&~ \mathcal{R}\times\frac{M_{n_-}}{4\pi M_{n_+}}\left(\frac{\text{Im}\left\{\kappa_{A+}\kappa^\ast_{A-}\left(U_{1,3}U_{3,2}|\kappa_{Bt}|^2+U_{3,3}^2\kappa_{B+} \kappa_{B-}^\ast \right) \right\}}{2|\kappa_{A-}|^2+|U_{3,2}^\ast\kappa^\ast_{Bt} |^2+|U_{3,3}^\ast\kappa_{B-}|^2}\right.
\nonumber\\
&+2\left.\frac{M_{t_{3B}}}{M_{n_+}} \frac{\text{Im}\left\{\kappa_{A+}\kappa^\ast_{A-}\left[U^\ast_{3,3}U^\ast_{3,2}\kappa_{B+}\kappa_{Bt}+U_{3,2}U_{3,3}\kappa_{B-}^\ast\kappa_{Bt}^\ast  \right] \right\}}{2|\kappa_{A-}|^2+|U_{3,2}^\ast\kappa^\ast_{Bt} |^2+|U_{3,3}^\ast\kappa_{B-}|^2}\right)~.\label{eq:epsm}
\end{align}
In these equations, $\mathcal{R}$ is a resonant factor for the intermediate $n$'s in the diagrams of Fig.~\ref{fig:asym} going nearly on-shell. In the limit $\Delta M\equiv M_{n_+}- M_{n_-} \sim \Gamma_N \ll M_N$, it is given by~\cite{Anisimov:2005hr}
\begin{align}
\mathcal{R}=&\frac{M_{n_+}}{M_{n_-}}\frac{M_{n_+} M_{n_-}(M_{n_+}^2-M_{n_-}^2)}{(M_{n_+}^2-M_{n_-}^2)^2+(M_{n_+}\Gamma_{n_+}-M_{n_-}\Gamma_{n_-})^2}~.
\label{eq:resnc}
\end{align}

When the mass eigenstates are far apart the resonant factor $\mathcal{R}$ approaches one and the asymmetry generation is not enhanced. Interestingly, the asymmetry generation is also suppressed when the $N$ masses are degenerate in the UV, $\xi\to 0$ in Eq.~\eqref{eq:Nmasses}. This can be seen by noticing that the imaginary part of the combination of the $\kappa$ couplings that appear in the numerator of Eqs.~\eqref{eq:epsp} and \eqref{eq:epsm} approaches zero in the $\xi\to0$ limit. In particular, when $\xi=0$, the off-diagonal elements of the matrix $M_N$ of Eq.~\eqref{eq:Nmasses} can be shown to be equal to those combinations of the $\kappa$ couplings. But since Eqs.~\eqref{eq:epsp} and~\eqref{eq:epsm}  are written in the mass eigenbasis the off-diagonal elements vanish. 

Consquently, a small but nonzero value for $\xi$ is optimal for the generation of the asymmetry. More precisely, the asymmetry generation is enhanced when the mass splitting of $n_{\pm}$ is small, but it becomes suppressed when the UV mass splitting in Eq.~\eqref{eq:Nmasses} (the term proportional to $\xi$) becomes smaller than the IR mass splitting (the term proportional to $c_{\Delta}$). In what follows we present numerical results for the size of the asymmetry for several values of $\xi$.

Finally, even when the $\epsilon_{A\pm}$ are sufficiently large, we still need to make sure that the asymmetry, once generated, is not washed out by subsequent processes. Since we have taken the reheat temperature to be low, processes mediated by an intermediate $\phi_A$ are inefficient. A different process that can reduce the asymmetry in the two sectors is the decay of $\phi_A$ to a visible quark and the twin RTC-singlet top. Now the partial width of $\phi_A$ decaying to visible states scales like $\lambda^2$ while the partial width to a visible quark and $t_{3B}$ scales like $\kappa_{At}^2$. Therefore the asymmetry washes out for $\kappa_{At} \gg \lambda$. Quantitatively, the washout factor of Eq.~\eqref{eq:defineeta} is
\beq
W=\frac{\Gamma(\phi_A\to \overline{D}_A+\overline{D}_A)}{\Gamma_{\phi_A}}~.
\eeq

\begin{figure}
	\begin{center}
		\includegraphics[width=0.49\textwidth]{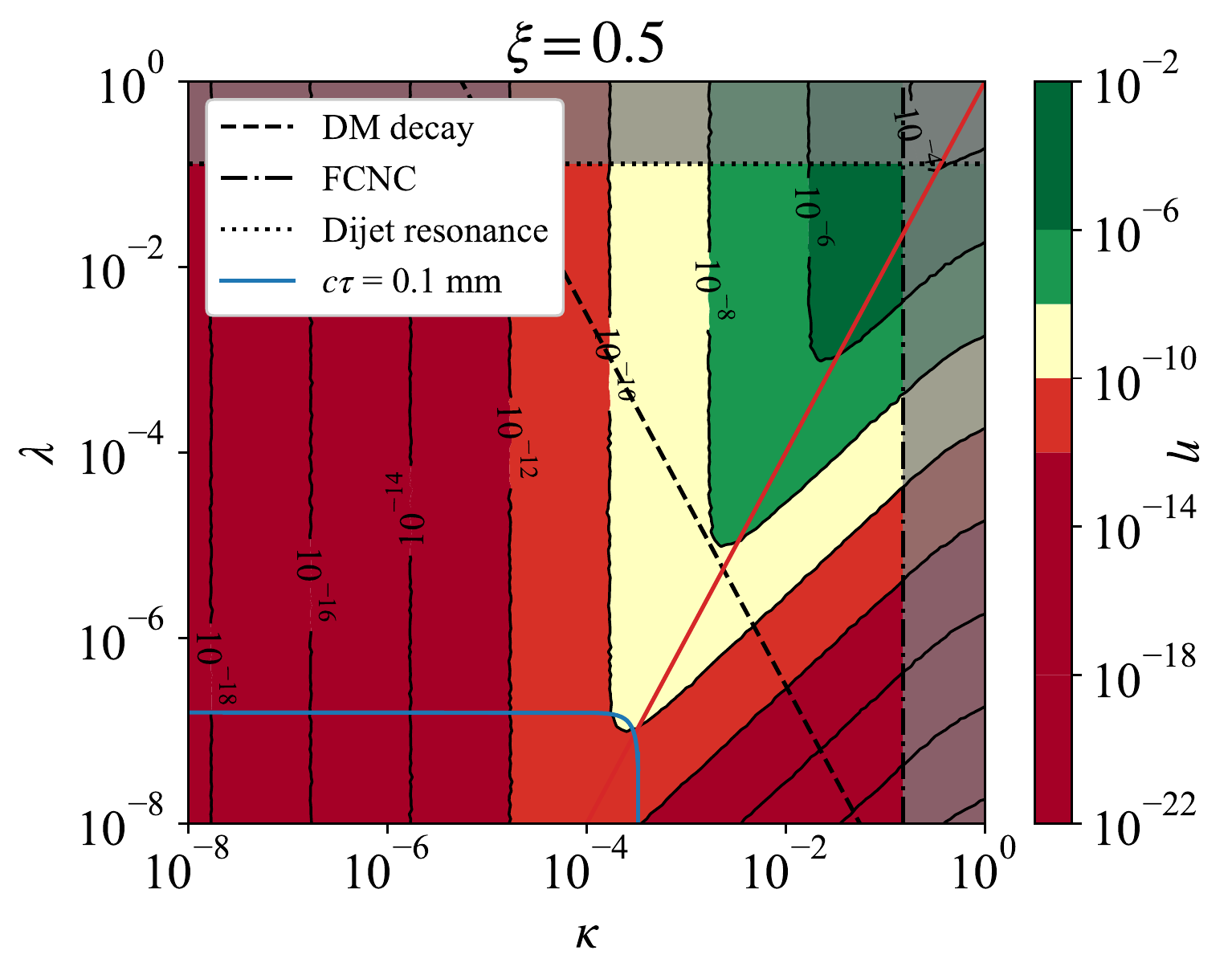}
		\includegraphics[width=0.49\textwidth]{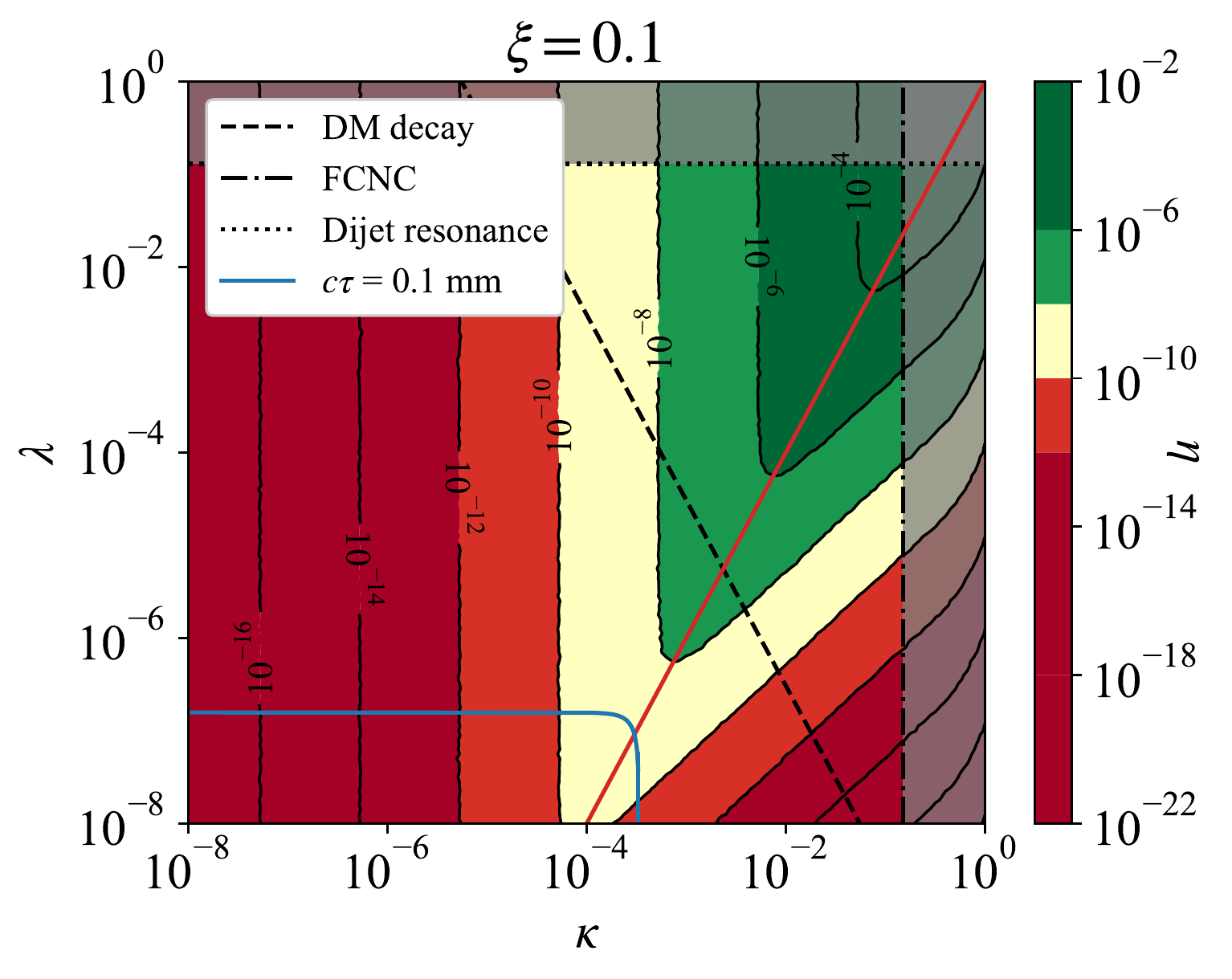}
		\includegraphics[width=0.49\textwidth]{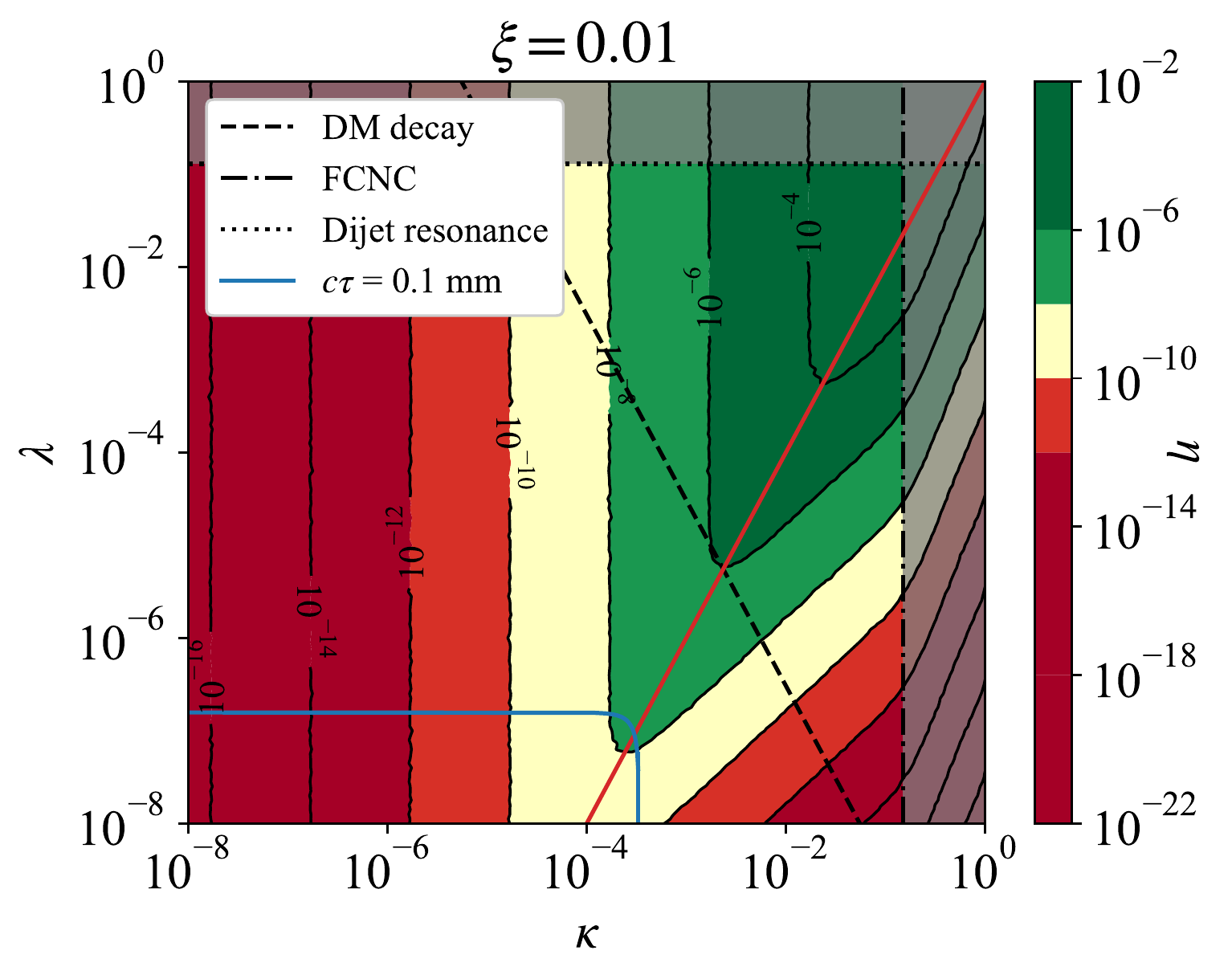}
		\includegraphics[width=0.49\textwidth]{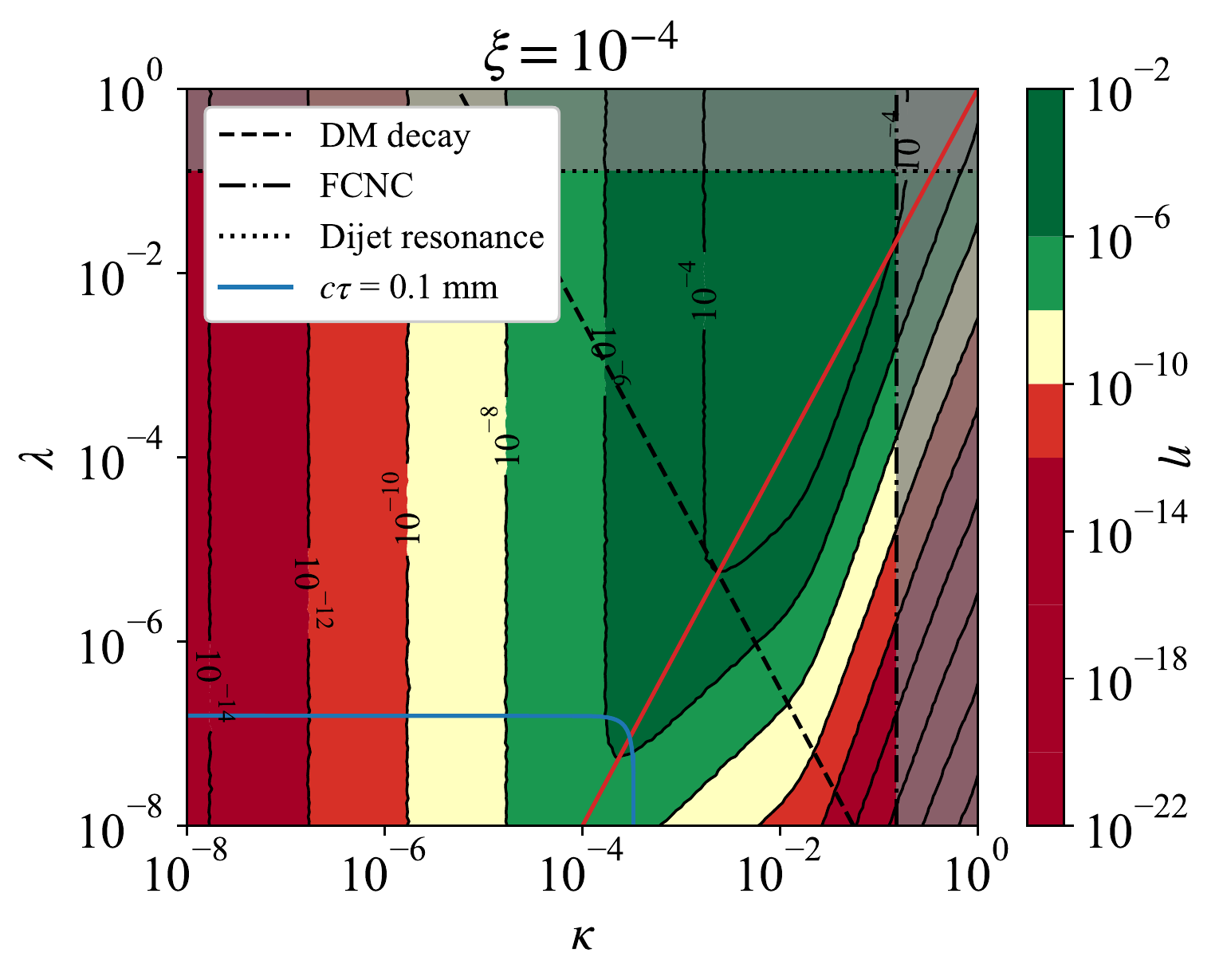}
		\caption{Contour plots for the median value of the asymmetry parameter $\eta$ as a result of the MC process described in the main text, as a function of $\kappa$ and $\lambda$, for several benchmark values of $\xi$. We take $m_\phi = 2$~TeV, $f_\phi=4$~TeV, and $m_N=4$~TeV. The vertical lines on the right correspond to FCNC constraints, see Sec.~\ref{sec:Precision}. The red diagonal lines show the boundary between the regions where $\phi_A$ decays dominantly to $jj$ (above) or $j/t+$MET (below). To the left and below the blue curve $\phi_A$ decays start becoming displaced, see Sec.~\ref{sec:collider}. The gray regions on top are the bounds on $\phi_A$ production from dijet resonance searches at the LHC. Finally, when $b_{3B}$ is kinematically allowed to decay, the region to the left of the diagonal dashed line is consistent with decaying DM constraints (this is based on the very conservative estimate of Sec.~\ref{sec:decayingDM}; the allowed region is likely larger. It is also possible that $b_{3B}$ is stable due to kinematics, in which case this constraint is entirely absent.)}		
		\label{fig:mainresults}
	\end{center}
\end{figure}

Having described the main parameter dependences in the generation of the asymmetry, we are ready to present our numerical results. As mentioned in Sec.~\ref{sec:simplified}, in calculating the asymmetry we cannot simply take all $\kappa$ entries to be equal to each other\textemdash among other things, there is no physical CP-violating phase in that case. Instead, we perform a Monte-Carlo based analysis. In Fig.~\ref{fig:mainresults}, for each point in the $\kappa$-$\lambda$ plane, we numerically calculate the asymmetry a large number of times. In each iteration, we randomly assign a magnitude to each of the (complex) $\kappa_{A,iI}$ and $\kappa_{B,I}$ elements in the interval $[0.5\kappa,2\kappa]$, and a random phase. We then calculate $\epsilon_{A\pm}$ and $W$, and we plot the median value of the resulting $\eta$ distribution using benchmark values of $m_\phi = 2$~TeV, $f_\phi=4$~TeV and $m_N=4$~TeV. The gray-shaded areas in these plots are ruled out due to phenomenological constraints, which are discussed in Sec.~\ref{sec:signatures}.

These plots exhibit the qualitative features already discussed. Smaller values of $\xi$ result in a larger resonant factor and a larger parameter space region that produces an acceptable value of $\eta$ (areas shaded green). In the bottom right corner of the plots the washout factor becomes important, and $\eta$ becomes smaller. In the red shaded areas, $\eta$ is too small to account for the observed baryon number in the universe. We take $10^{-8}$ to be the smallest phenomenologically acceptable value of $\eta$, since $Y_B = \eta Y_N$ and $Y_N \sim T_\mathrm{r} / M_\mathrm{r}$ \cite{Reece:2015lch}, $T_r$ being the reheat temperature, $\mathcal{O}(10~{\rm GeV})$, and $M_r$ being the mass of the reheaton, which must necessarily be above $M_N\gsim$~TeV.

%%%%%%%%%%%%%%%%%%%%%%%%%%%
\section{Signatures and Constraints}
\label{sec:signatures}

\subsection{Dark Photon constraints}
\label{sec:darkphoton}

 \begin{figure}
	\begin{center}
		\includegraphics[width=0.65\textwidth]{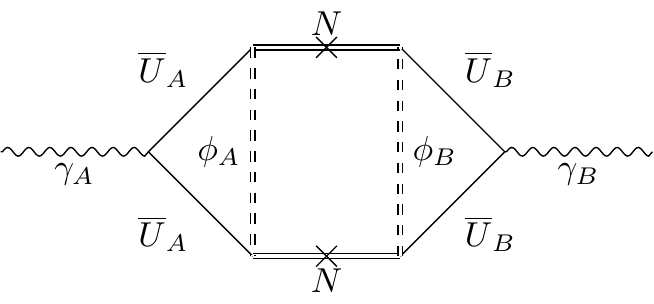}
		\caption{Feynman diagram for the contribution of the portal states to the kinetic mixing of the visible and twin photons. }
		\label{fig:kmx}
	\end{center}
\end{figure}

In Twin Higgs models, there is typically a small amount of kinetic mixing between the twin photon and the visible photon. Generically, the mixing is induced at four-loop level due to a mixing between the Higgs and the twin Higgs, and is of order $10^{-11}$~\cite{Koren:2019iuv}. In our model, the portal fermions also contribute to the kinetic mixing, as shown in Fig.~\ref{fig:kmx}. This mixing is estimated to be 
\beq
\varepsilon_{\rm portal} \sim \frac{e^2 \kappa^4}{(16\pi^2)^3} \approx 2.5 \times 10^{-8} {\kappa}^4.
\eeq
For $\kappa\sim 1$ this can be the dominant source of mixing. As mentioned in Sec.~\ref{sec:twin}, we can also include an explicit kinetic mixing term $B^{\mu\nu}B_{\mu\nu}'$ between the visible and twin hypercharge gauge groups, and consider $\varepsilon$ as an effectively free parameter, but not to be taken smaller than the dominant loop contribution.

There are a number of constraints on the kinetic mixing of dark photons, which are summarized in refs.~\cite{Fabbrichesi:2020wbt,Caputo:2021eaa}. These typically lead to upper limits on $\varepsilon$. In our model, there is also a lower bound on $\varepsilon$, which comes from demanding that when twin charged particles such as the twin tau (the symmetric component) efficiently annihilate to twin photons in the early universe, with the twin photons decaying sufficiently rapidly to SM particles. This limit can be expressed as~\cite{Pospelov:2007mp}
\beq
\Gamma_{\gamma'\to {\rm SM}}\gsim H(\tau_B\ {\rm freezeout}) \approx \frac{1}{M_{Pl}}\left(\frac{m_{\tau_B}}{20}\right)^2.
\eeq
For $m_{\gamma'}=1$~GeV, this translates to $\varepsilon_\text{min} \sim 5 \times 10^{-9}$. Since $m_{\tau_B}$ is of order GeV in our model, the twin photon cannot be heavier than that. While the upper bounds on $\varepsilon$ for $m_{\gamma'}=1$~GeV are only around $10^{-3}$, for lighter twin photons the bounds are significantly stronger. Therefore, we choose $m_{\gamma'} \sim 1$~GeV for our study, and consider $\varepsilon$ in the range $[10^{-8},10^{-3}]$. Mixing within these limits is consistent with the constraints from existing searches. Furthermore, parts of this $\varepsilon$ range are discoverable in ongoing experiments such as Belle II~\cite{Belle-II:2018jsg} as well as possible future experiments such as SHiP~\cite{SHiP:2015vad}.

\subsection{DM decay}
\label{sec:decayingDM}

As discussed in Sec.~\ref{sec:asymmetry}, we have $m_{b_{3B}}+m_{\tau_B}=5$~GeV. In this paper, we concentrate on the case where the twin bottom is the heavier of the two particles. We also take $m_{\tau_B}>1$~GeV so that the symmetric component of the twin taus can efficiently annihilate to twin photons, which have a $\sim$GeV mass for reasons mentioned in the previous section. The twin tau is exactly stable due to the unbroken twin lepton number. 

\begin{figure}[th]
	\begin{center}
		\includegraphics[width=0.5\textwidth]{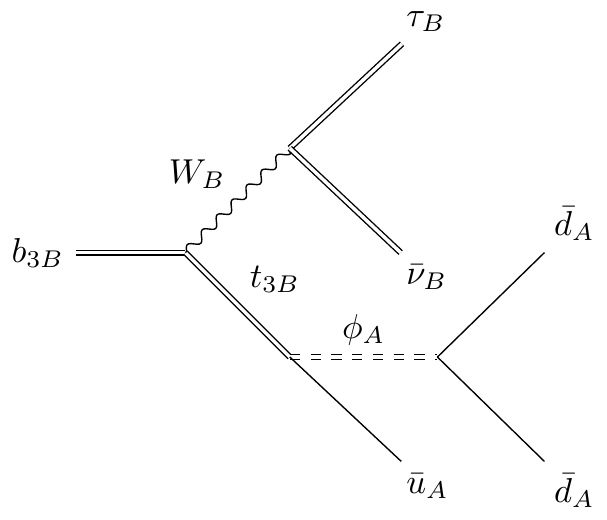}
		\caption{Feynman diagram for the twin bottom decay.}
		\label{fig:chiqqq}
	\end{center}
\end{figure}

As for the twin bottom, since $B_A-B_{B}'$ is conserved, and since there are no lighter twin states with nonzero baryon number, any potential decay mode must contain a SM antibaryon in the final state. For this decay to be kinematically allowed, the condition $m_{b_{3B}} > m_{\tau_B}+m_p$ has to be satisfied. This means that if $2.5~{\rm GeV}<m_{b_{3B}}<3~{\rm GeV}$, then the twin bottom is exactly stable, whereas in the range $3~{\rm GeV}<m_{b_{3B}}<4~{\rm GeV}$, the twin bottom can decay via the channel shown in Fig.~\ref{fig:chiqqq}.

The twin bottom decay proceeds through the $N$-portal, via an off-shell $W_B$, $t_{3B}$ and $\phi_A$. Note also that due to the antisymmetric flavor structure of the $\phi_A$ coupling, quark mixing via the CKM matrix needs to be involved in order for the final state quarks to hadronize into an antineutron. Being conservative and leaving out any hadronic form factors, we can parametrically put an upper bound on the width as follows:
\beq
\Gamma_{b_{3B}\to \bar{n}+{\rm invisible}} < \frac{m_{b_{3B}}^{11}}{8\pi(16\pi^2)^4} \frac{g_W^4}{m_{W_B}^4} \frac{f_\phi^2}{M_N^2} \frac{\kappa^4 \lambda^2}{m_{t_{3B}}^2 m_\phi^4}~,
\eeq
where we have taken into account the off-shell propagators, the 5-body phase space suppression, the mixing angle $f_{\phi} / M_N$ between $t_{3B}$ and the portal fermions, and the couplings in the diagram. While there are no dedicated constraints for the minimal decay mode DM$\to\bar{n}+$invisible, in order to be conservative we consider the possibility of other mesons being emitted in the decay, so we compare to decaying DM constraints into a generic hadronic final state ($qq$) at a mass of $m_{b_{3B}}-m_{\tau_B}-m_{\bar{n}}<2$~GeV (the maximum energy available for mesons in the final state), where the bound on the lifetime is $5\times10^{27}$~seconds \cite{Blanco:2018esa,Ackermann:2014usa}. The resulting constraint on the parameter space is shown as the diagonal dashed line in the panels of Fig.~\ref{fig:mainresults}.

\subsection{Precision observables}
\label{sec:Precision}

\begin{figure}
	\begin{center}
		\includegraphics[width=1\textwidth]{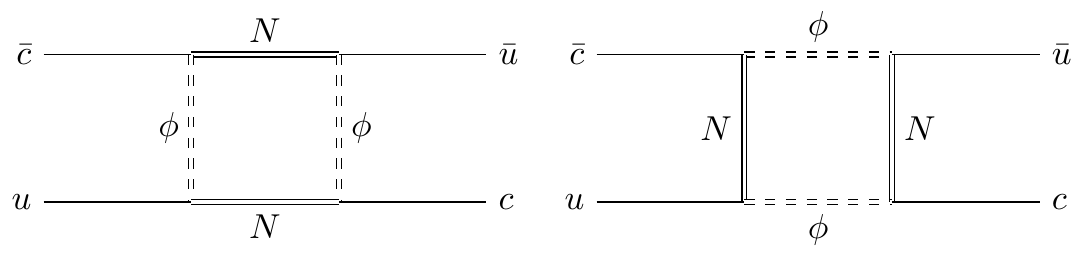}
		\caption{Leading Feynman diagrams contributing to $D^0 - \bar D^0$ mixing.}
		\label{fig:box}
	\end{center}
\end{figure}

We next turn our attention to constraints on flavor changing neutral currents (FCNC's), arising from Feynman diagrams such as those in Fig.~\ref{fig:box}. These induce charm meson mixing processes via effective operators such as
\beq
\mathcal L_\mathrm{FCNC} \supset - \tilde C^{uc} ( \bar{c} \bar \sigma_\mu u ) (  \bar{u} \bar \sigma^\mu c )  + \text{H.c.}~,
\label{eq:fcnceff}
\eeq
with coefficients of the form
\beq
\tilde C^{uc} \simeq \frac{\kappa_{A,1 \bar I} \kappa_{A,2  I}^* \kappa_{A,2 \bar J} \kappa_{A,1 J}^* }{8\pi^2 M_{N}^2}~,
\label{eq:fcnccp}
\eeq
for the diagram on the left in Fig.~\ref{fig:box}, and a similar expression for the diagram on the right, with the appropriate rearranging of indices.

The strongest constraints on FCNC processes come from $D^0 - \bar D^0$ mixing, which in our simplified coupling scheme gives $\kappa \lesssim \mathcal O(0.1)$. This is shown as the vertical line on the right in the panels of Fig.~\ref{fig:mainresults}.

 \begin{figure}
	\begin{center}
		\includegraphics[width=0.5\textwidth]{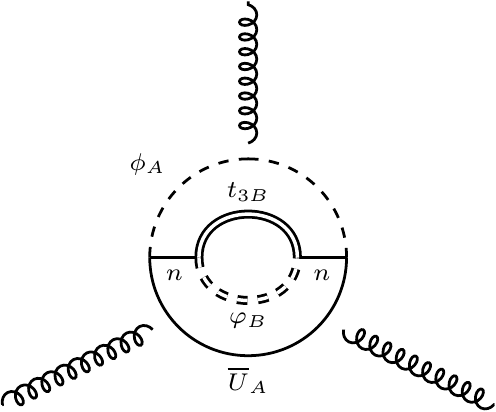}
		\caption{Feynman diagram for a contribution to the neutron EDM. The gluon lines can be attached in any possible way to SM colored particles.}
		\label{fig:edm}
	\end{center}
\end{figure}

Another potential observable is the generation of electric dipole moments (EDM's) due to the CP-violating phases in $\kappa$ couplings. Since the portal couplings involve quarks but not leptons, the main effect is a contribution to the neutron electric dipole moment. In effective field theory, this can be considered as a contribution to the Weinberg operator \cite{Weinberg:1989dx, Abe:2017sam, Hisano:2018bpz}
\beq
\mathcal L_\mathrm{\cancel{CP}} = - \frac{1}{3} \tilde C_G f^{ABC} e^{\mu\nu\rho\sigma} G^A_{\mu\lambda} G^{B\lambda}_\nu G^C_{\rho\sigma}~.
\eeq
A Feynman diagram contributing to this operator is shown in Fig.~\ref{fig:edm}. We estimate the size of the diagram parametrically as
\beq
\frac{3g_s^3}{(16\pi^2)^3}\frac{\kappa^4}{M_N^2}~.
\eeq
Comparing this estimate with the current best measured limit on the neutron EDM $d_n = (0.0 \pm 1.1) \times 10^{-26}~e\cdot \rm cm$ \cite{nEDM:2020crw} gives $\kappa \lesssim \mathcal O(1)$, not significantly constraining the parameter space. We see, however, that the new physics contribution can exceed the theoretical expectation in the SM of $|d_n| \sim 10^{-31}~e\cdot \rm cm$ \cite{Dar:2000tn}. Therefore, improvements in the measurement of $d_n$ as well as future proton EDM measurements~\cite{Lehrach:2012eg} can be sensitive to our model.

\subsection{Direct detection}

 \begin{figure}[H]
	\begin{center}
		\includegraphics[width=0.5\textwidth]{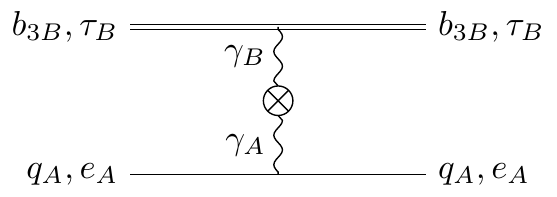}
		\caption{Direct detection contribution for $b_{3B}$ and $\tau_B$ through the photon/twin photon mixing.}
		\label{fig:fdd}
	\end{center}
\end{figure}

In our model, both $b_{3B}$ and $\tau_B$ can scatter off of nucleons due to the kinetic mixing between the visible photon and the twin photon, as shown in Fig.~\ref{fig:fdd}. These particles also have a contribution to nucleon scattering from the Higgs / twin Higgs mixing. However, the twin Higgs couplings to $b_{3B}$ and $\tau_B$ are Yukawa suppressed and the Higgs coupling to nucleons is only induced at loop level through the effective $h$-$g$-$g$ coupling, so this contribution is subdominant. 

The direct detection cross section in this channel was calculated in ref.~\cite{Essig:2017kqs}. Due to the low mass of the twin bottom and tau the most stringent constraints come from electron scattering, not nuclear scattering. We evaluate these constraints for our model, using the fact that both $b_{3B}$ and $\tau_B$ carry unit charge under the twin photon. Since the mass range of interest is relatively narrow, instead of a continuous scan we consider the endpoints of the range of interest, namely ($m_{b_{3B}}=2.5$~GeV, $m_{\tau_B}=2.5$~GeV) and ($m_{b_{3B}}=4$~GeV, $m_{\tau_B}=1$~GeV). We find that existing constraints are automatically satisfied for the entire range for $\varepsilon<10^{-3}$, which we assume due to other twin photon constraints as mentioned in Sec.~\ref{sec:darkphoton}. On the other hand, projecting to a future exposure of $10^5$~kg~yr for electron scattering experiments, the sensitivity region extends down to $\varepsilon\sim 3\times 10^{-4}$ for $m_{b_{3B}}=2.5$~GeV, and to $\varepsilon\sim 6\times 10^{-4}$ for $m_{b_{3B}}=4$~GeV. Therefore, future direct detection experiments will provide a valuable probe to the parameter space of our model.

\subsection{Collider phenomenology}
\label{sec:collider}

The relevant states for collider phenomenology in our model are $\phi_A$, $t_{3B}$, and portal fermions $n_{\pm}$. The goal of this paper is to present the model and the constraints on it from existing searches. While we also describe promising future directions for discovery, we do so in a relatively minimal way. We leave to future work more detailed studies of dedicated collider searches for the full range of possible production and decay channels.

\begin{figure}
	\begin{center}
		\includegraphics[width=0.49\textwidth]{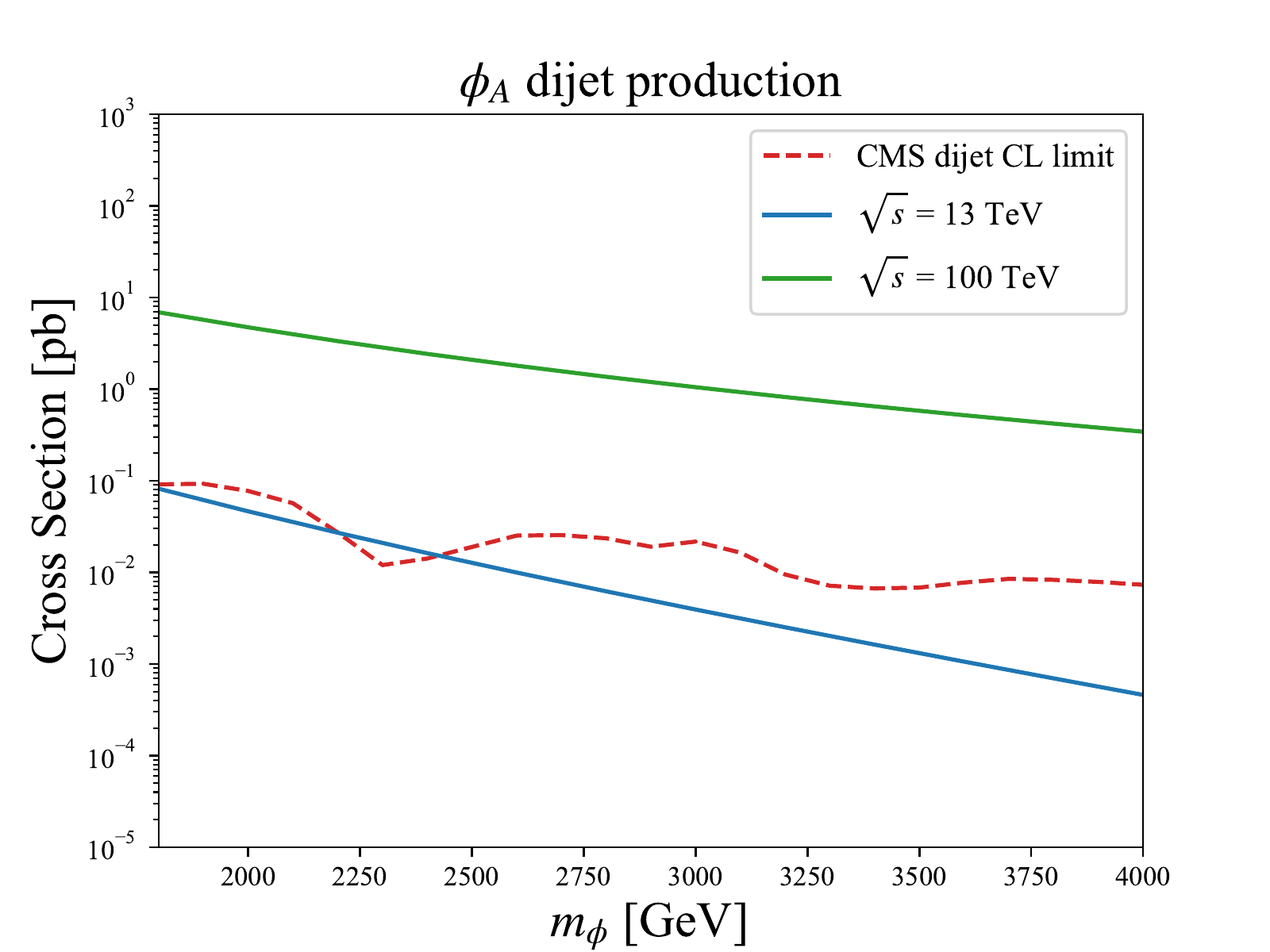}
		\includegraphics[width=0.49\textwidth]{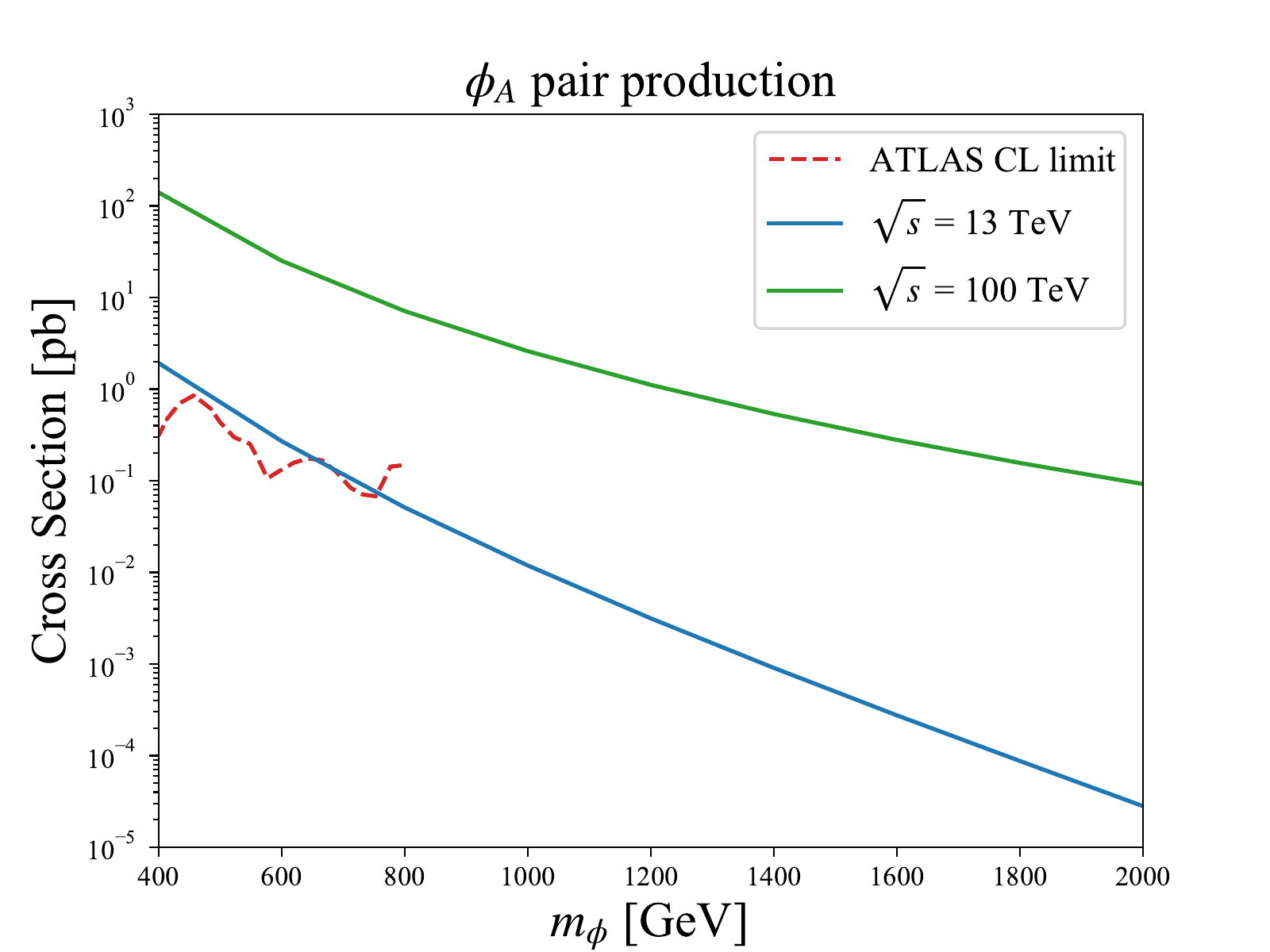}
		\caption{Left: Resonant production cross section for $\phi_A$ at the LHC and a future $p$-$p$ collider from a $d$-$s$ initial state, for $\lambda=0.1$, and the CMS dijet resonance bound for comparison~\cite{CMS:2019gwf}. Right: Pair production cross section for $\phi_A$ at the LHC and a future $p$-$p$ collider and the bounds from the ATLAS and CMS searches for $pp\to 2X\to 4j$ in the multijet final state~\cite{Sirunyan:2018rlj, Aaboud:2017nmi}. No branching ratios or cut efficiencies have been applied to the signal curves.}
		\label{fig:phiproduction}
	\end{center}
\end{figure}

As a color triplet, $\phi_A$ can be easily pair produced from a two gluon initial state. It can also be singly produced from a $d$-$s$ initial state via the $\phi$-$\overline{D}$-$\overline{D}$ interaction. In the left and right panels of Fig.~\ref{fig:phiproduction} we plot these production cross sections and the relevant experimental limits as a function of $m_{\phi_A}$. For the resonant production plot on the left, we use $\lambda=0.1$ for the signal cross section (with no branching ratios or cut efficiencies applied), and we plot the CMS bound on dijet resonances for comparison~\cite{CMS:2019gwf}. In the pair production plot on the right, we plot the ATLAS bound in the multijet final state for $pp\to 2X\to 4j$~\cite{Sirunyan:2018rlj, Aaboud:2017nmi} (again, no branching ratios or cut efficiencies have been applied to the signal).

The dominant decay mode of $\phi_A$ depends on its couplings. In the limit $\lambda\gg\kappa$, the dominant decay mode is two jets via the $\phi$-$\overline{D}$-$\overline{D}$ interaction, whereas in the opposite limit, $\phi_A$ can decay through the portal coupling to an up type quark and $t_{3B}$, which translates to $j/t+$invisible for practical purposes. Note that, since we assume $m_{\phi}<m_N$, this decay channel requires mixing between $t_{3B}$ and the $n_{\pm}$. Also, since we assume all $\kappa$ couplings to have similar sizes, the fraction of light jets and tops in this decay channel are comparable. In Fig.~\ref{fig:mainresults} we indicate with the red diagonal line the boundary between the two types of $\phi_A$ decays. In this figure, we also show with the blue curve in the bottom left, the region where $\phi_A$ decays start becoming displaced ($c\tau=0.1$~mm). The searches for the displaced decays~\cite{Sirunyan:2019gut,Lee:2018pag,Sirunyan:2020cao} are only sensitive up to $m_{\phi_A}\sim 1.8$~TeV. In this work we use $m_{\phi}=2$~TeV as a benchmark and we leave a dedicated analysis for lighter $\phi_A$ searches through displaced vertices to future work.

For sizable $\lambda$, the resonant production of $\phi_A$ is abundant, and dijet resonance searches provide a nontrivial constraint, restricting $\lambda\lsim 0.1$ for $m_{\phi_A}=2$~TeV. This is shown as the gray-shaded area on top of the panels in Fig.~\ref{fig:mainresults}. In comparison, the monotop final state is not significant in resonant production, because producing a large cross section through the $\lambda$ coupling requires suppressing the branching fraction into that decay mode. Dijet resonance searches will, of course, have greatly enhanced sensitivity at future colliders, see for instance~\cite{Helsens:2019bfw}.

The pair production cross section of $\phi_A$ is independent of the $\lambda$ and $\kappa$ couplings. When the dijet decay channel is the dominant one, existing bounds from multijet searches do not significantly constrain our model, as seen in the right panel of figure~\ref{fig:phiproduction}. When the $j/t+$invisible decay channel dominates, the final state signature is the same as for a pair produced RPV stop or sbottom, decaying to $t+$invisible. The reach for this production and decay mode has been projected for the HL-LHC~\cite{CidVidal:2018eel}, and for a 100~TeV hadron collider~\cite{FCC:2018vvp}. While our mass benchmark of $m_{t_{3B}}\sim 1$~TeV and $m_{\phi_A}\sim 2$~TeV will likely remain out of reach even at the HL-LHC, it is projected to be well within the reach of the 100~TeV hadron collider. This channel will likely be the most promising one in searches for $t_{3B}$. Kinematic observables such as $m_{T2}$ \cite{Lester:1999tx,Barr:2003rg,Cho:2007qv,Cho:2007dh,Cheng:2008hk,Burns:2008va} could then be used to determine the twin top mass.

\begin{figure}[th]
	\begin{center}
		\includegraphics[width=0.95\textwidth]{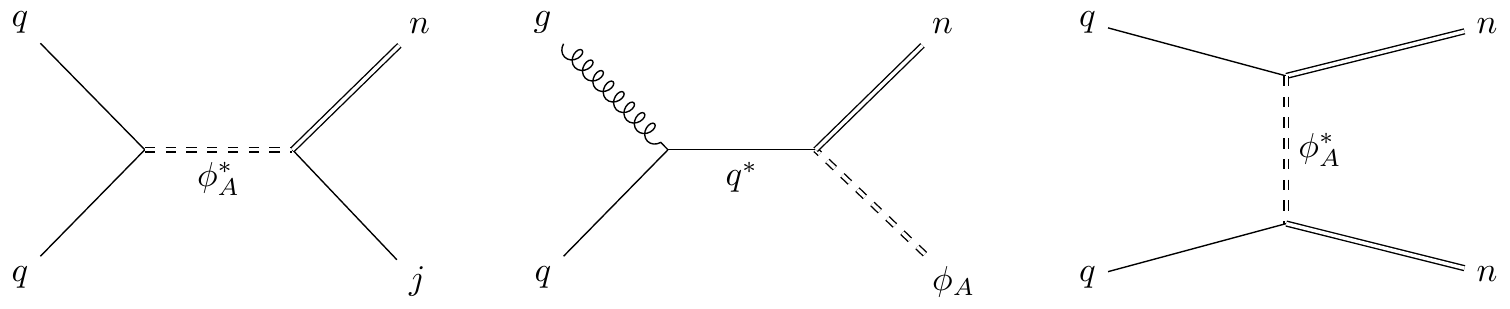}
		\caption{Production channels for the portal fermions.}
		\label{fig:nprod}
	\end{center}
\end{figure}

\begin{figure}[th]
	\begin{center}
		\includegraphics[width=0.65\textwidth]{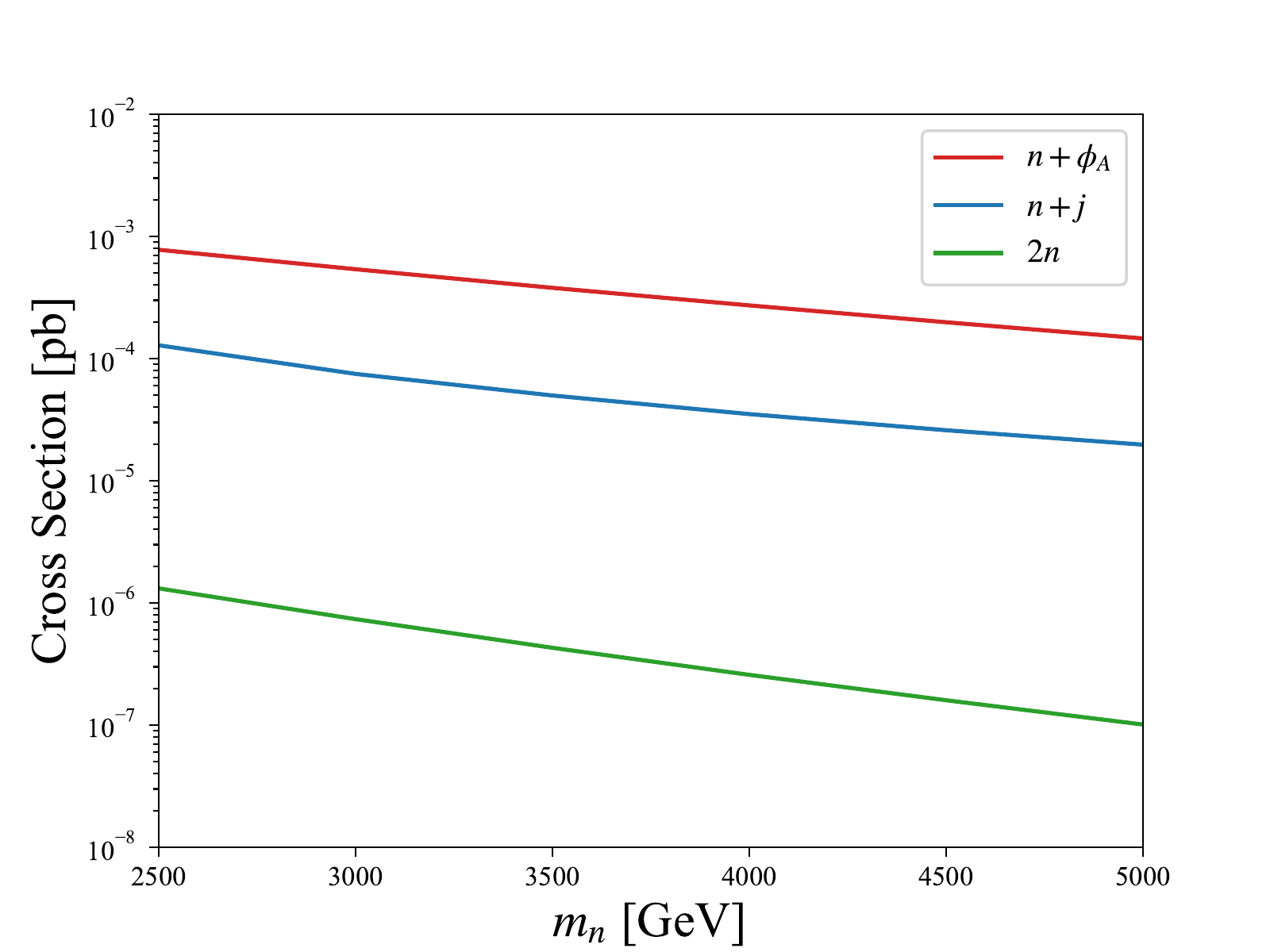}
		\caption{Cross section for different production channels shown in Fig.~\ref{fig:nprod} at $\sqrt s = 100 \tev$.}
		\label{fig:nscs}
	\end{center}
\end{figure}

The production channels for the portal fermions $n_{\pm}$ are shown in Fig.~\ref{fig:nprod}. These include single production recoiling against a jet or a top ($d\,\overline{s}\to\phi_A^\ast\to \overline{n}_\pm \overline{U}^\dag$), associated production with a $\phi_A$ ($\overline{U}g\to \phi_A\overline{n}_\pm$ ), and pair production through t-channel $\phi_A$ exchange ($\overline{U}^\dag\overline{U}\to \overline{n}_\pm\overline{n}^\dag_\pm$). Since the portal fermion masses are quite high, none of these production modes will be observable at the LHC, but in Fig.~\ref{fig:nscs} we plot the cross section of each at a 100~TeV hadron collider (for one of the $n_{\pm}$ in each case\textemdash for example the pair production cross section should be multiplied by four to include all possible final state combinations).

Once produced, the portal fermions decay either invisibly in the twin sector (to three twin tops, if that is kinematically allowed) or to $\phi_A$+$j/t$. This results in a variety of collider signals to search for and combine. In this work we limit ourselves to highlighting only one for plausibility of discovery, namely $n$-$t$ production, followed by $n\to \phi_A + t\to t\,\bar{t}+$MET. The final state thus contains three boosted tops, and missing energy. This rather distinctive final state can be identified using boosted top tagging techniques \cite{Kaplan:2008ie}, for a review containing additional references see \cite{Kogler:2018hem}. The leading backgrounds are expected to be $4t+(Z\to\nu\bar{\nu})$, and $t\,\bar{t}+(Z\to\nu\bar{\nu})+$jets with one hard jet being misidentified as a third top. A preliminary parton level study shows that with $p_T$ and MET cuts of order a TeV, discovery level statistical significance could be obtained in this channel even with conservative estimates for branching ratios and detector efficiencies. In future work, we will study this channel, as well as others relevant for the discovery of portal fermions, in rigorous quantitative detail.

\subsection{Exploring connections}
\label{sec:confirm}

In this section we have examined a wide range of possible signatures arising from our model. Here we comment on what can be learned from the connections between them. In particular, whether sufficient evidence can be gathered to point to the underlying physics addressing the naturalness, DM and M/AM asymmetry puzzles. It has been discussed in earlier work~\cite{Chacko:2017xpd,Kilic:2018sew,Chacko:2019jgi} that there may be sufficiently many potential measurements in the Higgs sector (both on properties of the SM-like state $h$ as well as the heavier state $H$) to make a strong case for the Twin Higgs mechanism as the answer to the naturalness puzzle.

In addition to those, in our model the HL-LHC or a future Higgs factory may provide sufficient precision to the SM-like Higgs invisible branching ratio that can be cross-checked against the existence of a twin bottom quark and a twin tau, with masses consistent with the condition $m_{b_B}+m_{\tau_B}=5$~GeV, providing a hint that the solutions of the DM and M/AM asymmetry puzzles may be connected.

Potential future experiments such as SHiP are expected to have broad sensitivity to the twin photon parameter space $(m_{A'},\varepsilon)$. If these parameters can be measured, then for $\varepsilon\gsim 10^{-6}$, direct detection experiments via electron scattering may provide a secondary probe into the existence of the twin bottom and the twin tau. If astrophysical parameters (such as the local DM velocity distribution) can be determined with sufficient precision, a statistical fit may then be able to confirm the existence of two DM components with different masses but with the same number density, further strengthening the case for a connection between the solutions to the DM and M/AM asymmetry puzzles.

Future increase in sensitivity in FCNC and EDM searches may provide evidence for beyond-the-SM contributions in both quantities, which in our model correspond to a best-fit region in the $(m_N,\kappa)$ parameters. Especially a discovery of EDM's beyond SM expectations will be a strong indication of new couplings with CP-violating phases, suggesting a connection with the M/AM asymmetry puzzle.

Finally, as we discussed in the previous section, searches at the HL-LHC and future collider experiments will have sensitivity to discover $\phi_A$, and potentially the twin top and portal fermions as well. The measurement of the masses ($m_{\phi_A}$, $m_{T_B}$, $m_{n_\pm}$) and couplings ($\lambda$, $\kappa$) of these particles would provide further cross-checks with the same parameters probed by other measurements (such as EDM searches), strengthening the case for a common mechanism underlying these phenomena.

%%%%%%%%%%%%%%%%%%%%%%%%%%%%%%%%%%%%%%%%%%%%%%%%%%%
\section{Conclusions}
\label{sec:conclusions}

We have presented a model where the fraternal Twin Higgs scenario is extended by adding colored scalars in the visible and twin sectors, and gauge singlet fermions that provide a new portal between the two sectors. This relatively modest addition results in rich phenomenological consequences. The portal fermions can initiate the reheating process and their decays can generate baryon asymmetries in the two sectors, the twin colored scalars can acquire a VEV and spontaneously break the twin color group down to an $SU(2)$ subgroup. As a result of this breaking, quarks that are singlets under the unbroken color group become DM candidates. Thus, the model in question can address the naturalness puzzle, the M/AM asymmetry puzzle and the DM puzzle. Furthermore, the $Z_2$ breaking mass term essential to having the twin and visible Higgs acquiring different VEVs is also generated as a result of twin color breaking.

There are large portions of parameter space where this model successfully addresses the puzzles in question, while remaining consistent with all existing experimental constraints for dark photons, decaying DM, FCNC and EDM searches, DM direct detection experiments, and collider searches in a variety of channels. Furthermore, future improvements in some of these experiments will have sensitivity to the available parameter space of the model. Future searches for dark photons, EDM's, direct detection of DM, and collider searches may yield crucial information, which, when observed in combination with one another, could provide strong hints for connections between the naturalness, DM, and M/AM asymmetry puzzles. In this paper, we have focused our effort mainly to describing the intricacies of the model, and evaluating the impact of existing experimental results on the parameter space. We leave to future work the more detailed estimates of dedicated searches in future experiments to the most promising discovery channels.

%%%%%%%%%%%%%%%%%%%%%%%%%%%%%%%%%%%%%%%%%%%%%%%%%%% 
\section*{Acknowledgements}

We are grateful to Zackaria Chacko and Chang Sub Shin for their insightful comments. The research of CK and TY is supported by
the National Science Foundation Grant Number PHY-1914679. The work of CBV was supported in part by National Science Foundation Grant Number PHY-1915005 and Simons Investigator Award \#376204.

%\appendix

%%%%%%%%%%%%%%%%%%%%%%%%%%%%%%%%%%%%%%%%%%%%%%%%

\bibliographystyle{JHEP}
\bibliography{THDMBG}{}
\end{document}